\documentclass[pra, 10pt, twocolumn, tightenlines, longbibliography, aps,superscriptaddress,nofootinbib]{revtex4-2}
\usepackage{amsmath}
\usepackage{amssymb}
\usepackage{graphicx}
\usepackage{color}
\usepackage{soul}
\usepackage[utf8]{inputenc}
\usepackage[T1]{fontenc}
\usepackage{array}
\usepackage{comment}
\usepackage{multirow}
\usepackage[caption=false]{subfig}
\usepackage{hyperref}
\hypersetup{
	colorlinks=true,
	linkcolor=blue,
	filecolor=magenta,     
	urlcolor=blue,
    citecolor=blue,
	pdftitle={Overleaf Example},
	pdfpagemode=FullScreen,
}

\begin{document}

\title{Controlling the rain fall statistics using Mean-Reverting Jump Diffusion model}

\author{Joya GhoshDastider}
\affiliation{Department of Physics, Indian Institute of Technology, Guwahati 781039, Assam, India}

\author{D. Pal}
\affiliation{Department of Physics, Indian Institute of Technology, Guwahati 781039, Assam, India}

\author{Pankaj Kumar Mishra}
\affiliation{Department of Physics, Indian Institute of Technology, Guwahati 781039, Assam, India}

\date{\today}

\begin{abstract} 
We present a stochastic mean-reverting jump-diffusion model to simulate rainfall time series and validate it using long-term half-hourly rain fall data from the North-East region of India. The model captures the intermittent and extreme-event dynamics of rainfall, reproducing superdiffusive behavior with an exponent $\sim 1.8$, along with the observed probability distributions and multifractal features. By systematically varying key parameters, we demonstrate a transition between Log-Normal and Gamma distributions, and show how the occurrence of extreme events and dry-patch durations can be controlled. Spectral and wavelet analyses further confirm that the simulated series reproduces the dominant temporal scales observed in real rainfall data. Our proposed framework provides a robust tool for generating realistic synthetic rainfall series and serves as an effective approach for understanding the influence of underlying stochastic processes that governs the rainfall statistics.
\end{abstract}

\flushbottom

\maketitle

\section{Introduction}\label{sec1}

Rainfall is a highly complex phenomenon, influenced by a multitude of factors including temperature, humidity, air pressure, topography, and other atmospheric conditions~\cite{neelin2022precipitation}. The intensity of rainfall in a given region has significant implications for agricultural productivity, water resources, biodiversity, and even economic stability~\cite{hasan2010simple,arvind2017statistical,muthiah2024analyzing,duan2023application}. Consequently, accurately modeling rainfall events is crucial. However, due to the nonlinear and chaotic nature of rainfall, this remains a challenging task.

A closer examination of rainfall time series reveals that the occurrence of rainfall events is primarily governed by two interrelated dynamics. The first concerns the occurrence or non-occurrence of rainfall, while the second pertains to the stochastic nature of rainfall intensity during rainy periods~\cite{PhysRevLett.122.158702}. In this paper, we refer to the rainy periods as ``wet patches'' and the dry intervals as ``dry patches''~\cite{fall2021performance,tichavsky2019dry}. Notably, the occurrence of both wet and dry patches is inherently random.

To capture these two distinct processes, a combination of discrete and continuous models is required: a discrete process to model the onset of rain and a continuous process to simulate the fluctuating intensity of rainfall during wet patches. Several approaches have been proposed to model rainfall events. Among the most commonly used deterministic models are (i) the General Circulation Model (GCM)~\cite{mandal2019reservoir,gouda2018evaluation,yoo2018comparison,meher2017performance} and (ii) the Statistical Dynamical Model (SDM)~\cite{kurihara1970statistical,gowri2022assessment,abeysingha2023assessment,aieb2020statistical,demirdjian2018statistical}. The GCM, which is based on physical conservation laws (momentum, mass, energy, moisture, etc.), provides a detailed, albeit computationally intensive, simulation of rainfall patterns in specific regions. However, GCM’s accuracy and applicability are limited by computational costs and the lack of precise physical information~\cite{raju2020review}. Conversely, SDM relies on statistical averaging across space or time to estimate rainfall intensity, but this approach tends to oversimplify the system by neglecting inherent fluctuations, making it less suitable for capturing the full complexity of natural rainfall processes~\cite{vallis1982statistical}.

In recent years, researchers have turned to stochastic models to more effectively capture the intrinsic randomness of the rainfall process. A notable contribution was made by K. Hasselmann~\cite{hasselmann1976stochastic} in 1976, who introduced a stochastic term into climate models to account for internal fluctuations. This concept was further extended to investigate features of Sea Surface Temperature anomalies~\cite{frankignoul1977stochastic,lemke1977stochastic}.

Stochastic modeling~\cite{northrop2024stochastic} has since become a prominent approach in climate research, particularly in capturing the highly variable components of climatic systems. Numerous stochastic models have been proposed to simulate rainfall time series, including those based on the Markov Chain method. In the Markov Chain approach, each day’s rainfall state (wet or dry) depends on the previous day’s state, and the model can be extended to higher orders to incorporate the influence of multiple preceding days~\cite{ng2018generation,hassan2014application}. While Markov Chain models have been extensively used, their limitations arise in accurately simulating rainfall over larger temporal and spatial scales, particularly when it comes to low-frequency rainfall events. Additionally, specific parametric distributions such as Exponential, Gamma, Weibull, and Log-Normal have been used to model rainfall amounts in different regions~\cite{ng2018generation,li2014assessing,gao2020development}.

However, these models often require pre-estimated parameters, which can introduce biases in the probability distributions and underestimate extreme events. To address this issue, hybrid models incorporating Bayesian techniques have been developed, allowing for a probabilistic relationship between rainfall intensity and wet patch duration~\cite{zelalem2023bayesian}. While such models improve rainfall intensity prediction, they are still limited in their ability to predict future rainfall events. Other statistical models, like the Auto-regressive Integrated Moving Average (ARIMA), combine auto-regressive and moving average components to handle seasonal and non-seasonal patterns in rainfall data~\cite{qin2018estimating,wei2017potential,alam2024enhanced}. However, these models lack sufficient physical understanding of the rainfall time series, making them less suited for capturing the underlying dynamics.

Many of the aforementioned models typically separate the dynamics of rainfall occurrence and intensity, though these processes are inherently connected and continuous in time. To bridge this gap, researchers have introduced models like the Ornstein-Uhlenbeck (OU) process and its various modifications~\cite{bassanoni2025rare,eliazar2009ornstein}, which incorporate both the occurrence and intensity dynamics in a continuous framework. Recently, a censored power-transformed OU process was employed to model rainfall events, separating the seasonal and stochastic components while using a L\'evy subordinate process to account for extreme rainfall events~\cite{tong2021censored}. Although such models offer valuable insights, they do not fully capture the complete set of characteristics present in the rainfall process.

In this work, we develop a stochastic Jump Diffusion model~\cite{daly2006probabilistic,lefebvre2025first} to simulate rainfall time series, using data from the North-East region of India. This model is designed to capture the key features of the observed rainfall patterns in the region, including its statistical and spectral properties.

Our analysis of half-hourly rainfall data from the North-East region over the past twenty years reveals that rainfall intensity follows a multiplicative Log-Normal distribution. The spectral analysis of this data shows a power-law behavior with distinct exponents in the high and low-frequency regions ($1.5$ and $0.3$, respectively)~\cite{ghoshdastider2025kolmogorov}. Wavelet and Intrinsic Mode Decomposition (IMD) techniques reveal local time scales present in the rainfall time series, and multifractal analysis further quantifies the complexity of the rainfall process across different regions. The distribution of rainfall amplitudes varies according to topography, internal rainfall processes, and regional characteristics, with the Gamma and Log-Normal distributions being the most commonly observed~\cite{cho2004comparison,amburn2015precipitation,bhavana2012modeling}.

In this paper, we propose a minimal stochastic model for rainfall events using the mean-reverting Jump Diffusion process. We show that by adjusting the model parameters, we can obtain both Gamma and Log-Normal distributions for rainfall, and that the statistical and spectral characteristics of the simulated data closely match those observed in the North-East region of India.

The paper is organized as follows. In Sec.~\ref{sec2}, we discuss the stochastic features of real-world rainfall time series. In Sec.~\ref{sec3}, we introduce our Jump-Diffusion model for rainfall. In Sec.~\ref{sec4}, we present the results of our model simulations and compare them with observed rainfall data, highlighting their statistical, spectral, and multifractal characteristics, as well as the influence of varying parameters on the model’s behavior in different regions.

\section{Stochastic characteristics of rainfall time series}\label{sec2}

In general, rainfall time series exhibit two distinct yet interrelated patterns. 
The first is the periodic transition between non-raining (dry) and raining (wet) 
states, which reflects the intermittent nature of precipitation driven by 
atmospheric processes. The second is the stochastic variability in rainfall 
intensity during the wet periods, where the amount of rain can fluctuate 
significantly over short intervals. The interplay between these two dynamics, namely,
intermittency and intensity variability, renders rainfall time series a complex 
and fascinating phenomenon to analyze and model~\cite{ghoshdastider2025kolmogorov}.  

\begin{figure}[h!]
    \centering
        \includegraphics[width=0.48\textwidth]{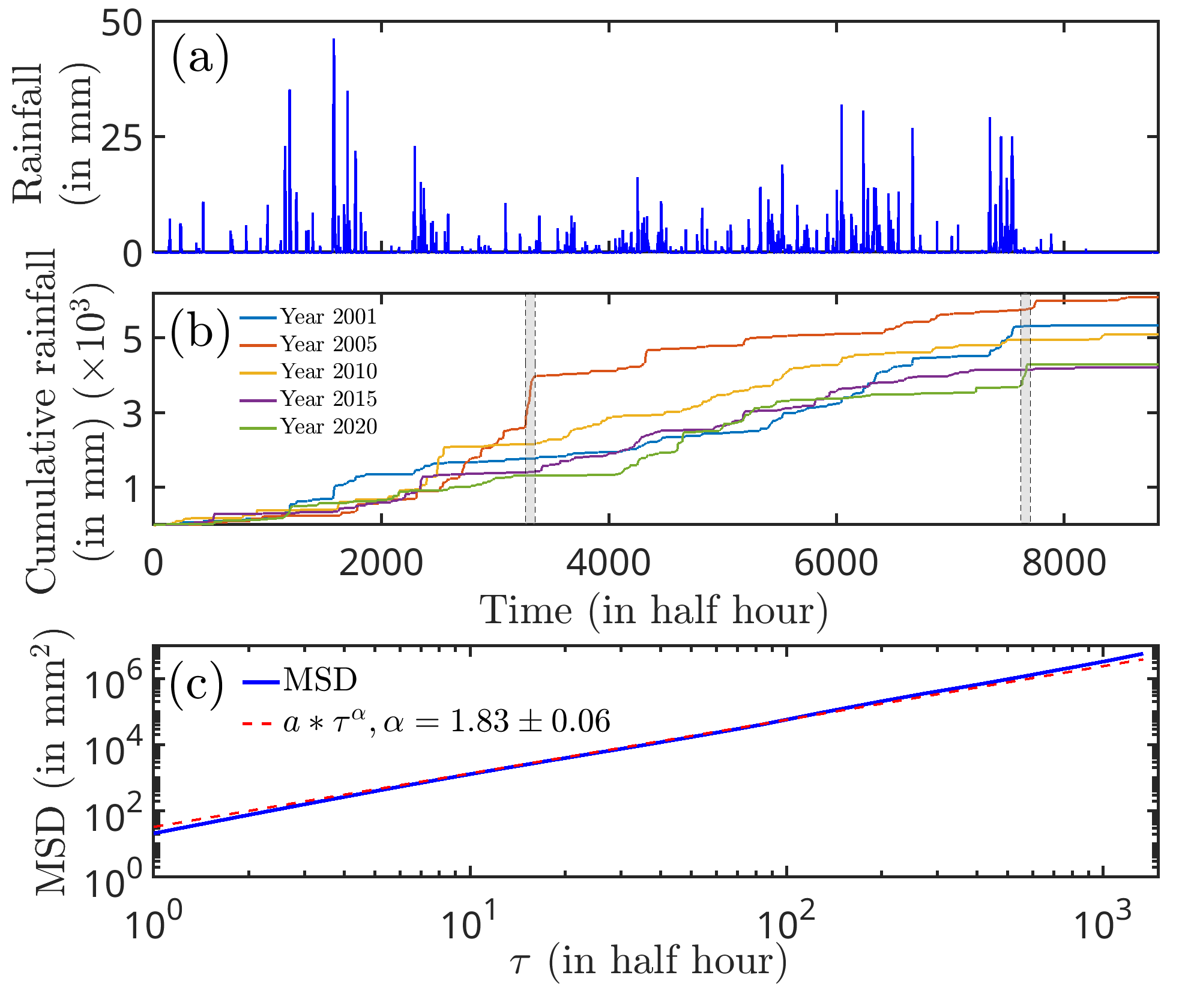}
    \caption{Temporal evolution of observed rainfall data of the station $26.05^{\circ}N,88.05^{\circ}E$ of year $2001$. (a) Rain fall Time series, (b) Cumulative time series of the same station for different years (indicated by different colors), where gray shaded area indicates the jumps due to extreme events, (c) Time averaged Mean Square Displacement (MSD) of the cumulative rainfall time series (blue) plotted against the time lag $\tau$ on a log-log scale. The dotted red line shows a power-law fit $\tau^{\alpha}$, with exponent $\alpha = 1.83 \pm 0.06 (>1)$, indicating superdiffusive dynamics.} \label{fig1}
\end{figure}

In this section, we present rainfall time series data collected from different 
measurement stations and characterize their statistical properties using 
stochastic tools such as cumulative rainfall and mean-square displacement (MSD). 
As an illustrative example, Fig.~\ref{fig1}(a) shows the temporal evolution of 
observed rainfall data recorded at the station located at $26.05^{\circ}N, 
88.05^{\circ}E$ during the year 2001. Alongside this, Fig.~\ref{fig1}(b) presents 
the cumulative rainfall corresponding to the same dataset, which provides a 
useful representation of the total rainfall accumulated over time.

The cumulative rainfall time series is defined as:
\begin{equation}
C_{t} = \sum_{i=1}^{t} X_{i},
\label{eq:ct}
\end{equation}
where $X_{i}$ denotes the rainfall amplitude at time step $i$, and $C_{t}$ 
represents the total accumulated rainfall up to time $t$. This transformation 
effectively converts the original rainfall signal into a trajectory that can be 
interpreted as a stochastic process.

Interestingly, the cumulative rainfall exhibits a diffusive-like behavior, 
characterized by alternating phases of plateaus and abrupt jumps. The plateau 
regions correspond to dry spells, during which the rainfall intensity remains 
below a predefined threshold (taken here as $0.001$ mm per half-hour). In 
contrast, the sharp upward jumps represent wet periods, including episodes of 
heavy or extreme rainfall. These jumps, highlighted by the gray shaded regions 
in Fig.~\ref{fig1}(b), reflect bursts of precipitation that significantly 
contribute to the overall accumulation.

To quantitatively analyze the diffusive nature of this cumulative process, we 
compute the mean square displacement (MSD), a widely used measure in stochastic 
process analysis. The time-averaged MSD is defined as~\cite{rosen2021mean, 
schirripa2024trajectory, he2008random}:
\begin{multline}
MSD (\tau = k \Delta t) = \frac{1}{N-k+1} \sum_{i=0}^{N-k} 
\left|X_{t_{i}+k} - X_{t_{i}}\right|^{2}, \\
k = 1, 2, ..., N-1,
\label{eq:msd}
\end{multline}
where $N$ is the total number of observations in the time series $X_{t}$, $k$ 
denotes the lag index, and $\Delta t$ is the temporal resolution of the data. 
The MSD essentially measures how the magnitude of fluctuations evolves with 
increasing time lag $\tau$.

Figure~\ref{fig1}(c) displays the MSD of the cumulative rainfall time series as 
a function of $\tau$ on a log-log scale (solid blue curve). The observed scaling 
behavior follows a power-law form, $MSD(\tau) \sim \tau^{\alpha}$, with a scaling 
exponent $\alpha = 1.83 \pm 0.06$, as indicated by the red dotted line. The value 
of $\alpha$ provides insight into the nature of the diffusion process: 
$\alpha = 1$ corresponds to normal diffusion, $\alpha < 1$ indicates 
subdiffusion, and $\alpha > 1$ signifies superdiffusion~\cite{alves2016characterization, 
rosen2021mean}. The exponent obtained here clearly points to superdiffusive 
behavior, suggesting the presence of long-range temporal correlations and the 
influence of extreme rainfall events. The turbulent transport of magneto-hydrodynamic thermal convection in the sun ($\alpha=1.57$)~\cite{rincon2025observational}, ultraquantum turbulence in both vortex segments ($\alpha=1.56$) and superfluid parcels ($\alpha=1.67$)~\cite{tang2025turbulent} show superdiffusive characteristic with similar exponent values.

These observations highlight essential statistical features that any realistic 
numerical model of rainfall must capture. In particular, both the intermittent 
occurrence of rainfall and the variability in its intensity need to be 
incorporated. In our modeling framework, the occurrence of rainfall events is 
assumed to be governed by the arrival of clouds, which we represent as a Poisson 
process. Each cloud arrival triggers a rainfall event, contributing to the 
underlying cumulative process and thereby introducing stochasticity and 
diffusive characteristics into the system. This combined mechanism provides a 
physically motivated and mathematically tractable approach to modeling the 
complex dynamics of rainfall time series.

\section{Stochastic Model of Rainfall Dynamics}\label{sec3}

In order to develop a comprehensive and physically meaningful model of rainfall dynamics, it is essential to first understand the underlying statistical properties of observed precipitation data. Rainfall is inherently a complex geophysical process influenced by a wide range of atmospheric factors, leading to significant variability across both space and time. Therefore, any realistic modeling approach must be grounded in empirical observations and supported by rigorous statistical analysis.

With this motivation, we begin by carefully examining the statistical characteristics of long-term observational data. Specifically, we analyze real-time rainfall records from the North-East region of India, covering the geographical span of \(26.05^{\circ}\text{N} - 26.95^{\circ}\text{N}\) and \(88.05^{\circ}\text{E} - 94.95^{\circ}\text{E}\), over a twenty-year period from 2001 to 2020~\cite{ghoshdastider2025kolmogorov}. This region is climatologically significant, as it includes some of the wettest places on Earth.

A detailed statistical investigation reveals several important features of the rainfall dynamics. Among these, one of the most crucial observations is that the rainfall intensity follows a Log-Normal probability distribution function~\cite{ghoshdastider2025kolmogorov}. This implies that while moderate rainfall events are most common, there exists a non-negligible probability of extremely high-intensity rainfall events. Such extreme events manifest as a fat tail in the probability distribution, which is a characteristic signature of this region’s rainfall pattern. This heavy-tailed behavior reflects the frequent occurrence of intense rainfall bursts, consistent with the region’s reputation for experiencing some of the highest rainfall levels globally~\cite{kuttippurath2021observed}.

Motivated by these empirical findings, we model the rainfall dynamics using a mean-reverting stochastic process with jumps, specifically an Ornstein-Uhlenbeck-type stochastic differential equation augmented with a jump component~\cite{nafidi2019stochastic}. The governing equation is given by
\begin{equation}
dP_t = k(\mu - P_t)\,dt + \sigma P_t\, dW_t + J_t\, dN_t,
\label{eq1}
\end{equation}
where $P_t$ represents the rainfall intensity at time $t$. The parameter $\mu$ denotes the long-term mean rainfall level, while $\sigma$ characterizes the magnitude of fluctuations. The parameter $k > 0$ is the mean-reversion rate, ensuring that the process tends to relax toward the mean value over time. The term $W_t$ denotes a standard Wiener process, capturing the continuous stochastic variability in rainfall. Different variants of this model are  used in several non-linear systems in the study of risk-return characteristics of pair trading in finance, Dynamics of population growth etc~\cite{stubinger2018pairs,nafidi2019stochastic}.

To incorporate the sudden and intense rainfall bursts observed in the data, we include a jump component $J_t dN_t$. Here, $N_t$ is a Poisson process with intensity $\lambda$, representing the random arrival of rainfall-generating events such as cloud bursts. The variable $J_t$ denotes the amplitude of these jumps and is assumed to follow a log-normal distribution, consistent with the empirical distribution of rainfall intensity. The probability of a single jump occurring in a small time interval $(t, t+dt)$ is $\lambda dt$, while the probability of multiple jumps within this interval is negligible.

We further refine the model by considering the physical interpretation of a jump event. If the rainfall intensity at time $t$ is $P_t$, then after a jump it changes multiplicatively to $J_t P_t$. Therefore, the net change due to a single jump is $(J_t - 1)P_t$. Incorporating this into Eq.~\eqref{eq1}, we obtain
\begin{equation}
dP_t = k(\mu - P_t)\,dt + \sigma P_t\, dW_t + (J_t - 1)P_t\, dN_t.
\label{eq2}
\end{equation}

Since $J_t$ is log-normally distributed, it follows that $(J_t - 1)$ is also log-normally distributed. Let its mean and standard deviation be denoted by $\theta$ and $\sigma_1$, respectively. Taking expectations, we obtain
\begin{equation}
\mathbb{E}[(J_t - 1)dN_t] = \mathbb{E}[J_t - 1] \, \mathbb{E}[dN_t] = \theta \lambda dt,
\label{eq3}
\end{equation}
where $\mathbb{E}[\cdot]$ denotes the expectation operator. This shows that the jump component contributes a predictable drift term to the process.

However, for a proper stochastic representation, it is desirable that the jump contribution remains purely random and does not introduce a deterministic bias. To ensure this, we compensate for the mean contribution of the jump term by adjusting the drift component. Specifically, we subtract $\theta \lambda P_t dt$ from the drift, leading to the modified equation
\begin{equation}
dP_t = \big(k\mu - kP_t - \theta \lambda P_t\big)dt + \sigma P_t\, dW_t + J_t\, dN_t.
\label{eq4}
\end{equation}

This final form represents a mean-reverting stochastic process with multiplicative noise and random jumps, capturing both the continuous fluctuations and intermittent extreme events observed in rainfall data.

To study the behavior of this model, we numerically solve Eq.~\eqref{eq1} using the Euler--Maruyama scheme~\cite{PhysRevE.70.017701,yu2024convergence}, which is a standard method for approximating solutions of stochastic differential equations. The details of this numerical implementation are provided in Appendix~\ref{app1}. The simulated data obtained from this scheme allows us to compare the model predictions with observed rainfall statistics and validate the effectiveness of our approach. For this purpose, we set different parametric values to simulate rainfall time series that closely resembles the observed rainfall data: mean-reversion rate ($k=0.15$), time step ($dt=1$), mean of the rainfall time series ($\mu=0.03$ mm), standard deviation of the rainfall time series ($\sigma=0.007$), mean of the amplitude of the arrival process ($\theta=0.8$), standard deviation of the arrival process ($\sigma_{1}=1$) and mean intensity of the arrival process ($\lambda=0.02$). In the later part of this paper, we explore the effect of different parametric values ($\mu, \sigma, \theta, \sigma_{1}, \lambda$) on the simulated rainfall data and their statistical, spectral and multifractal characteristic features. 

\section{Simulation Results}\label{sec4}

In this section, we present the numerical simulation results of the stochastic model (Eq.~(\ref{eq1})) to generate synthetic rainfall time series and employ that to investigate different statistical features to characterize the simulated data. Further we compare the simulated data with the observed rainfall data by extracting the required parameters by analyzing half-hourly real-time rainfall data of the North-East region ($26.05^{\circ}N-26.95^{\circ}N$, $88.05^{\circ}E-94.95^{\circ}E$) of India from May to October over twenty years ($2001-2020$). Using this stochastic model, we probe different parametric regions and investigate the behavior of simulated rainfall time series in those regions.  

\begin{figure}[h!]
    \centering
        \includegraphics[width=0.48\textwidth]{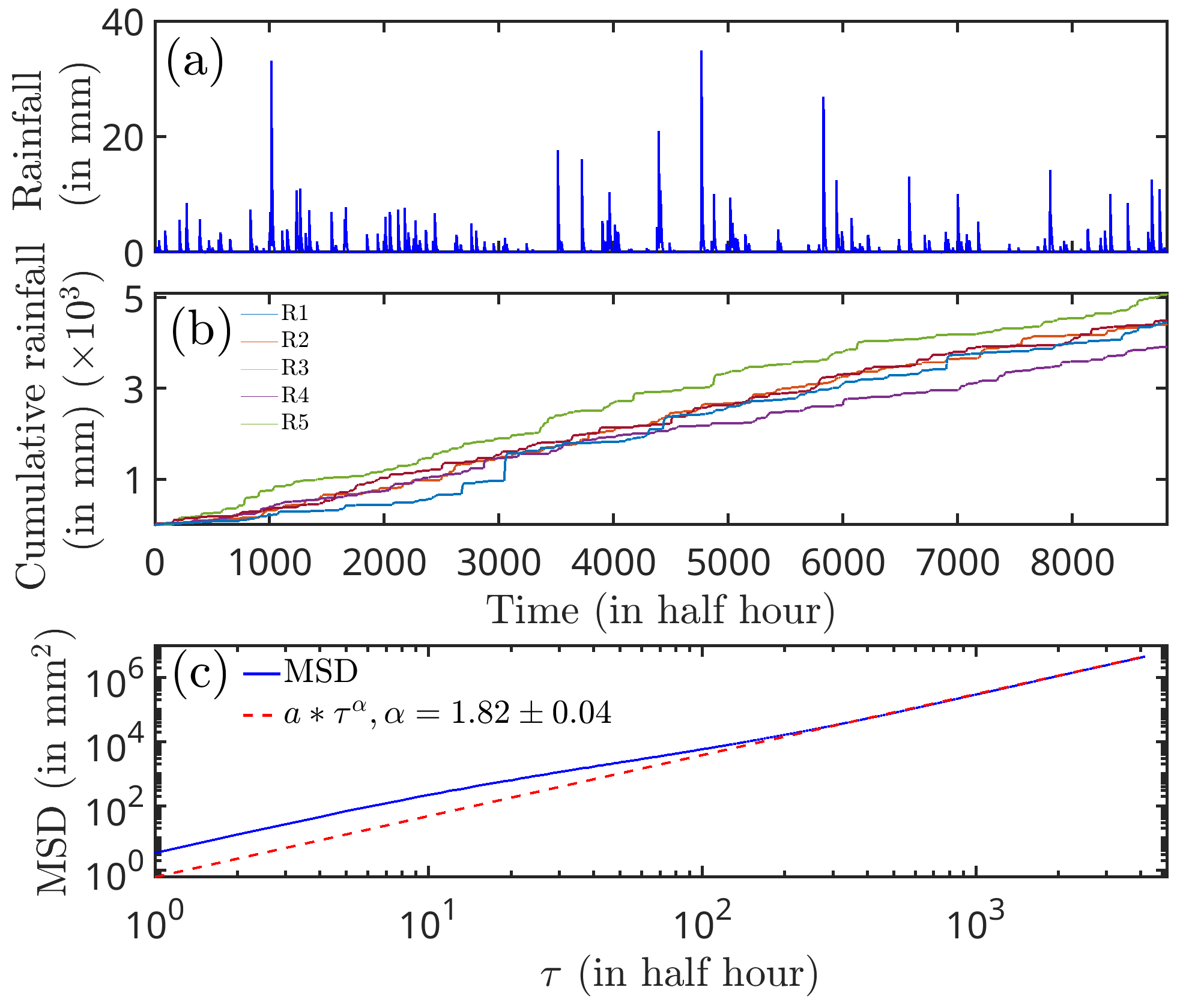}
    \caption{Temporal evolution of the simulated rainfall data. (a) Time series, (b) Cumulative time series for different realizations (indicated by different colors named as R1, R2 etc.), (c) Mean Square Displacement (MSD) of the simulated cumulative rainfall time series varies with time lag $\tau$ with an exponent $\alpha=1.82 \pm 0.04$ in log-log scale. $\alpha >1$ indicates that it is a superdiffusive process which matches with the observed result. }
    \label{fig2}
\end{figure}

In Fig.~\ref{fig2}, we present the temporal evolution of the rainfall time series obtained by numerically simulating Eq.~(\ref{eq1}). Panel (a) of Fig.~\ref{fig2} shows the simulated rainfall intensity as a function of time, while panel (b) depicts the corresponding cumulative rainfall time series, computed using Eq.~\ref{eq:ct}.

The simulated data successfully reproduce the characteristic intermittency observed in real rainfall records. In particular, the cumulative rainfall exhibits a stochastic diffusive trajectory marked by alternating plateau regions and sharp upward jumps. The plateaus correspond to dry intervals with negligible rainfall, whereas the abrupt jumps represent wet periods associated with intense precipitation events. These features closely resemble those observed in the empirical data shown earlier in Fig.~\ref{fig1}, indicating that the model effectively captures the intrinsic stochastic structure of rainfall dynamics.

To further quantify this behavior, we compute the time-averaged mean square displacement (MSD) of the simulated rainfall time series using Eq.~\ref{eq:msd}. The results are shown in Fig.~\ref{fig2}(c), where the MSD is plotted as a function of the time lag $\tau$ on a log-log scale (solid blue line). The MSD exhibits a clear power-law scaling with $\tau$, characterized by an exponent $\alpha = 1.82 \pm 0.04$, as indicated by the red dotted fitting line. Notably, this exponent is in close agreement with the value obtained from the observed rainfall data (Fig.~\ref{fig1}), thereby confirming that the simulated time series also exhibits superdiffusive behavior. This agreement demonstrates the ability of the proposed stochastic model to realistically capture both the qualitative and quantitative features of rainfall variability.

To make our analysis more general and robust, we extend our study by computing rainfall time series over a broad range of model parameters (see Appendix~\ref{app2} for a detailed discussion). In particular, Fig.~\ref{app_fig1} presents the simulated time series corresponding to different regions of the parameter space. Each row in this figure illustrates the effect of varying one of the five key parameters in Eq.~(\ref{eq1}). Specifically, the first row corresponds to changes in the mean rainfall level ($\mu$), the second row represents variations in the standard deviation of rainfall fluctuations ($\sigma$), the third row shows the effect of the mean jump amplitude ($\theta$), the fourth row captures the role of the standard deviation of the jump amplitude ($\sigma_{1}$), and the fifth row corresponds to variations in the intensity of the arrival process ($\lambda$).

The three columns in Fig.~\ref{app_fig1} represent different regimes of each parameter. The first column corresponds to lower values within the chosen parameter range, the second column represents intermediate values for which the statistical and spectral properties of the simulated rainfall closely match those of the observed data, and the third column shows the behavior for higher parameter values.

From Fig.~\ref{app_fig1}, we observe a systematic dependence of the simulated rainfall intensity on these parameters across their respective ranges. This parametric sensitivity is crucial for understanding how different aspects of the model influence the statistical behavior of rainfall time series. In particular, it provides insight into key features such as the probability distribution function, the occurrence of extreme events, and other higher-order statistical properties, which are discussed in detail in the later sections of the paper.

\begin{figure*}[!htp]
    \centering
        \includegraphics[width=1\textwidth]{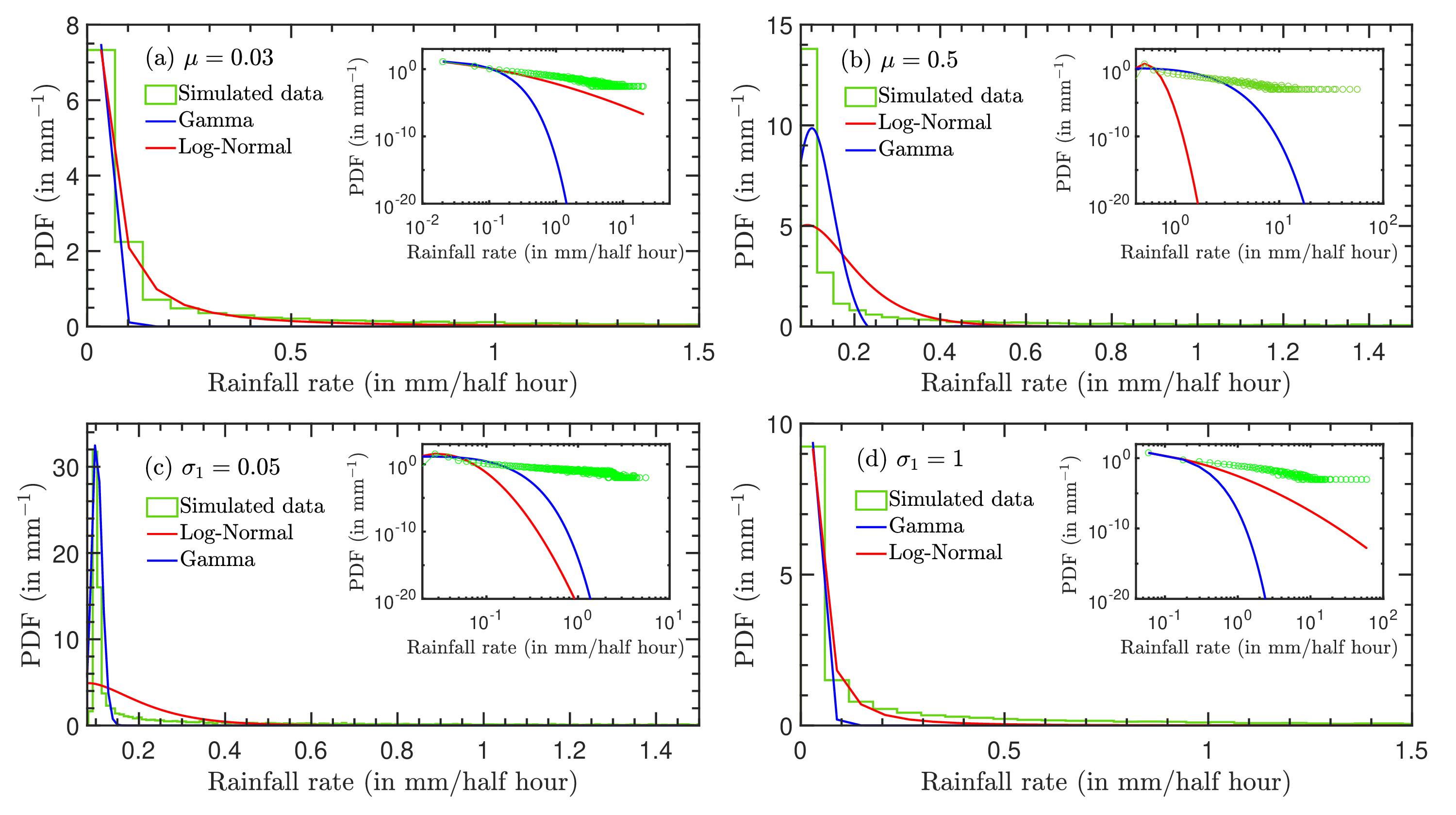}
    \caption{PDF of simulated rainfall data using Eq.~(\ref{eq1}). Green histograms represent the simulated rainfall amplitudes, while the red and blue curves show Log-Normal and Gamma PDF fits, respectively. Panels (a) and (b) illustrate a transition from Log-Normal to Gamma PDF as the arrival amplitude parameter $\theta$ increases from $0.03$ to $0.5$. Panels (c) and (d) show a shift from Gamma to Log-Normal PDF as the standard deviation of the arrival amplitude $\sigma_{1}$ increases from $0.05$ to $1$. Insets display the same PDFs on a log–log scale for better visualization of the distribution tails. }
    \label{fig3}
\end{figure*}

\subsection{Statistical analysis of simulated rainfall data}

Determining the statistical nature of rainfall distribution has long remained a challenging problem, primarily due to its strong dependence on multiple factors such as topography, intrinsic atmospheric processes, and regional climatic conditions. Over the years, numerous studies have attempted to characterize rainfall patterns across different parts of the world using a variety of statistical models. However, despite these efforts, a universally accepted distribution describing rainfall variability is still lacking. In practice, several candidate distributions—such as Gumbel, Pearson type III, Gamma, Log-Gamma, Normal, and Log-Normal (with two parameters)—have been employed to fit rainfall data~\cite{kurniawan2019distribution}. Among these, the Gamma and Log-Normal distributions are most frequently reported in the literature~\cite{cho2004comparison,amburn2015precipitation,bhavana2012modeling}.

It has been observed that rainfall time series from many regions of the world follow a Log-Normal probability distribution function (PDF)~\cite{cho2004comparison}, while in other regions, the Gamma distribution provides a better fit. This variability suggests that rainfall statistics are highly sensitive to underlying physical and environmental conditions. Motivated by this, in the present work we systematically investigate our stochastic model across a wide range of parameters and demonstrate how a transition from a Log-Normal distribution to a Gamma distribution can be achieved by appropriately tuning the model parameters.

In the following sections, we examine various statistical and dynamical features of the simulated rainfall time series to assess how well the model reproduces the characteristics of observed rainfall data, particularly from the North-East region of India. To this end, we compute several quantitative measures, including the probability density function (PDF), power spectrum, wavelet spectrum, and multifractal spectrum. These analyses enable us to identify key signatures of rainfall variability and to understand the influence of different model parameters on the emergent statistical behavior.

\subsubsection{Probability distribution function (PDF) of rain fall}
In Fig.~\ref{fig3}, we present the probability density function (PDF) of the simulated rainfall data in the form of a histogram (green), along with fitted Gamma (blue) and Log-Normal (red) distributions. The inset shows the same distributions plotted on a log-log scale to better visualize the tail behavior.

In the upper panel of Fig.~\ref{fig3} ((a) and (b)), we illustrate the PDF for two different values of the mean rainfall parameter ($\mu$), while keeping the other parameters fixed [$\sigma = 0.007, \theta = 0.03, \sigma_{1} = 1, \lambda = 0.02$]. In Fig.~\ref{fig3}(a), the Log-Normal distribution provides a better fit to the simulated data compared to the Gamma distribution. This observation is consistent with our earlier analysis of observed rainfall time series from the North-East region of India~\cite{ghoshdastider2025kolmogorov}, where the Log-Normal distribution was found to describe the rainfall statistics more accurately. In contrast, Fig.~\ref{fig3}(b) demonstrates that for a different value of $\mu$, the Gamma distribution offers a superior fit over the Log-Normal distribution. This highlights the crucial role of the mean rainfall parameter in determining the nature of the distribution.

We further observe that, similar to $\mu$, the standard deviation of the jump amplitude ($\sigma_{1}$) also significantly influences the form of the PDF. This is illustrated in the lower panel of Fig.~\ref{fig3} ((c) and (d)), where we show the PDFs for two different values of $\sigma_{1}$ while keeping the remaining parameters unchanged. In Fig.~\ref{fig3}(c), the Gamma distribution fits the simulated data more effectively, whereas in Fig.~\ref{fig3}(d), the Log-Normal distribution provides a better description. These results clearly demonstrate that both $\mu$ and $\sigma_{1}$ play a key role in governing the statistical nature of rainfall intensity.

\begin{figure}[!htp]
    \centering
        \includegraphics[width=0.5\textwidth]{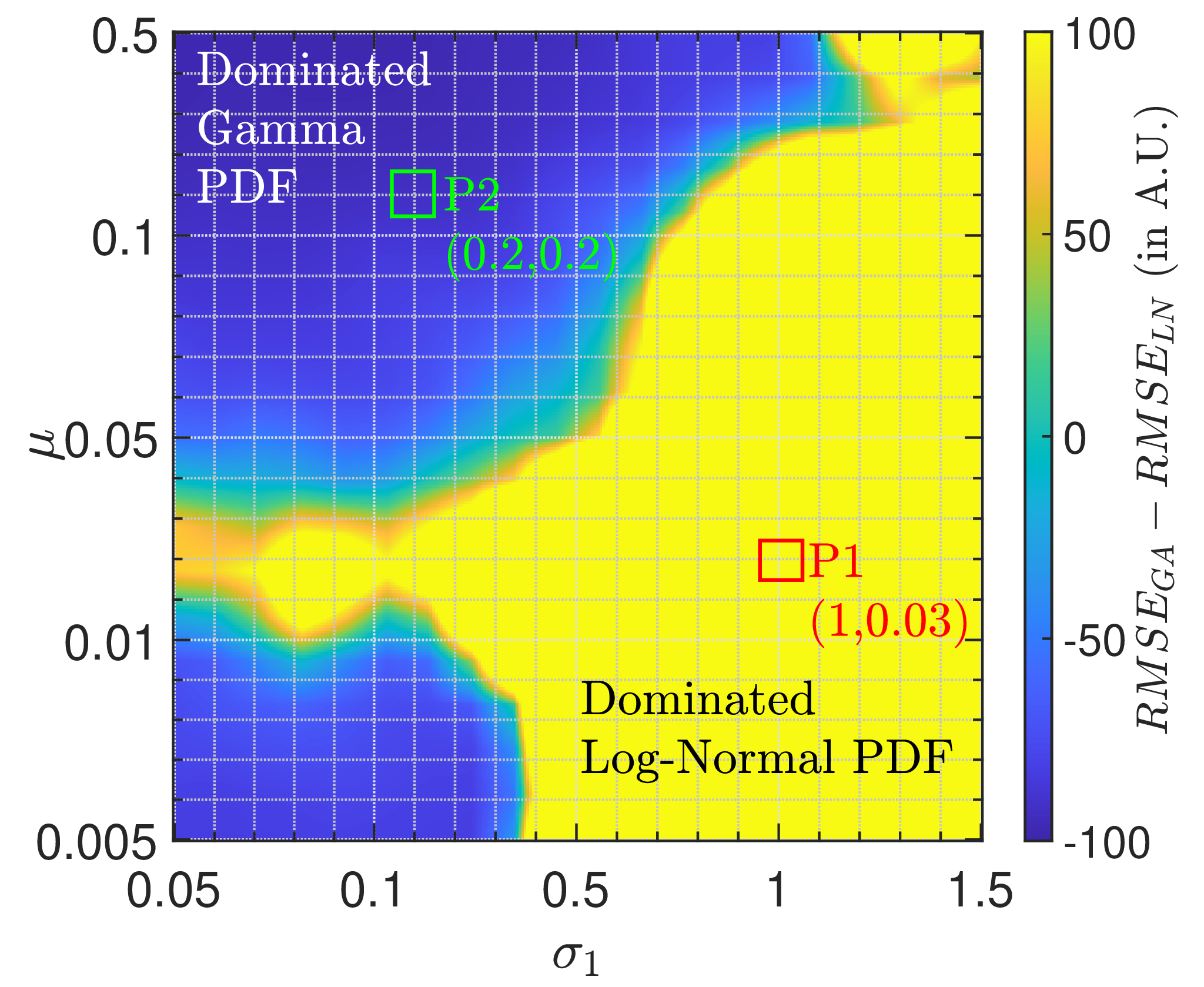}
    \caption{Phase diagram of PDF fitting for simulated rainfall data in $\mu-\sigma_1$ plane. The pseudo color map shows the difference in Root Mean Square Error (RMSE) between Gamma and Log-Normal PDF fits. Positive values indicate regions where the Log-Normal PDF provides a better fit, while negative values indicate regions dominated by the Gamma PDF. This diagram highlights how the underlying model parameters control the transition between the two characteristic rainfall distributions.}
    \label{fig4}
\end{figure}

To provide a more comprehensive representation of this behavior, we construct a phase diagram shown in Fig.~\ref{fig4}. In this figure, the x-axis represents different values of $\sigma_{1}$, while the y-axis corresponds to different values of $\mu$. The color bar indicates the difference between the root mean square errors (RMSE) of the Gamma and Log-Normal fits to the simulated rainfall data. The RMSE is computed using the relation~\cite{sagar2023encyclopedia}:
\begin{equation}
RMSE = \sqrt{\frac{1}{n} \sum_{i=1}^{n} |A_{i} - F_{i}|^{2}},
\label{eq5}
\end{equation}
where $A_i$ and $F_i$ denote the actual and fitted values, respectively.

In Fig.~\ref{fig4}, $RMSE_{GA}$ and $RMSE_{LN}$ denote the RMSE values corresponding to the Gamma and Log-Normal fits, respectively. A positive value of $(RMSE_{GA} - RMSE_{LN})$ indicates that the Log-Normal distribution provides a better fit, whereas a negative value implies that the Gamma distribution is more appropriate. The phase diagram clearly delineates regions dominated by Gamma fitting (dark blue) and Log-Normal fitting (yellow) in the $\mu$–$\sigma_{1}$ parameter space.

This analysis demonstrates that, through appropriate parameter tuning, the proposed stochastic model is capable of reproducing different statistical behaviors observed in rainfall data across various geographical regions. In Fig.~\ref{fig4}, we highlight two representative parameter sets: P1 (marked by a red rectangle), located in the Log-Normal dominated region, and P2 (marked by a green rectangle), situated in the Gamma dominated region. Notably, the parameter set P1 corresponds to conditions under which the model successfully reproduces the characteristic features of observed rainfall data from the North-East region of India.

Having established how the probability distribution of the simulated rainfall data varies with model parameters, we now proceed to analyze the spectral and multifractal properties of the system. For this purpose, we focus primarily on the two representative parameter sets, P1 ($\mu=0.03, \sigma=0.007, \theta=0.8, \sigma_{1}=1, \lambda=0.02$) and P2 ($\mu=0.2, \sigma=0.007, \theta=0.8, \sigma_{1}=0.2, \lambda=0.02$) , to gain deeper insight into the underlying dynamics.

 \subsubsection{Statistics of extreme events}

In this work, we consider that during the rainy phase, if the intensity of rain is equal to or greater than $13.5$ mm/half hour, then we call it an extreme event and it is in line with the guidelines set by Indian Meteorological Department~\cite{srivastava2016variability,bhattacharyya2022characteristics}. The presence extreme events in the rainfall time series suggests the existence of high convective flow which can potentially lead to floods, landslides, and other related natural calamities. In the present paper, we incorporate the impact of extreme event in our stochastic model through the stochastic jump which has been considered as having the random lognormal distribution. Among these parameters we are interested to analyze the impact of the fluctuation of the amplitude of the arrival process ($\sigma_{1}$) and the mean intensity of the arrival process ($\lambda$) which also in turn show great impact of the occurrence of extreme events. 

\begin{figure}[!htp]
    \centering
        \includegraphics[width=0.5\textwidth]{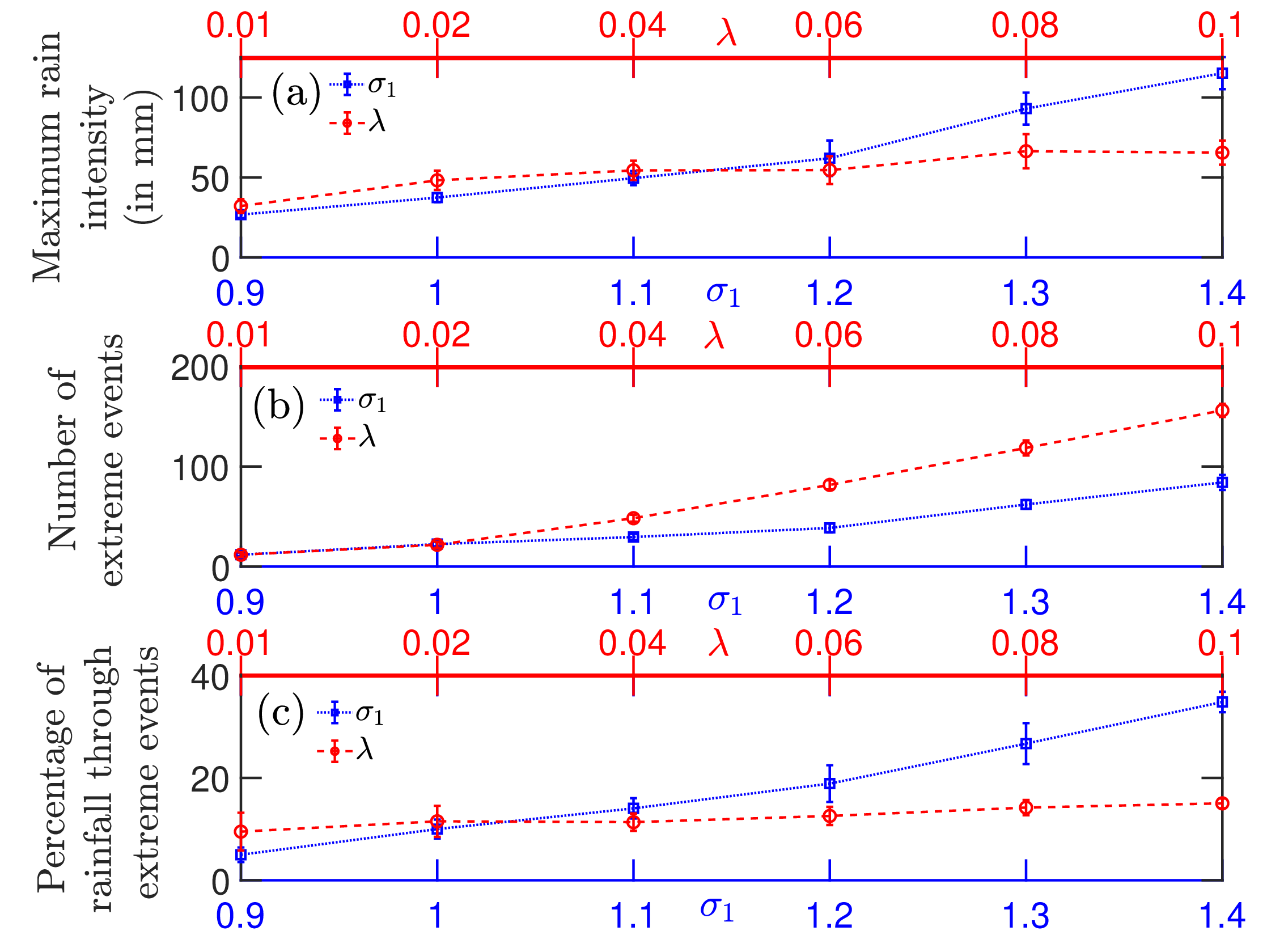}
    \caption{Behaviour of extreme events with parameter variation. (a) The variation of average highest amplitude of rain of simulated dataset with increasing values of $\lambda$ and $\sigma_{1}$. (b) The variation of number of extreme events in a dataset with $\lambda$ and $\sigma_{1}$. (c) The variation of percentage of rain through extreme events in simulated datasets with increasing $\lambda$ and $\sigma_{1}$.}
    \label{fig5}
\end{figure}

\begin{figure*}[!htp]
    \centering
        \includegraphics[width=1\textwidth]{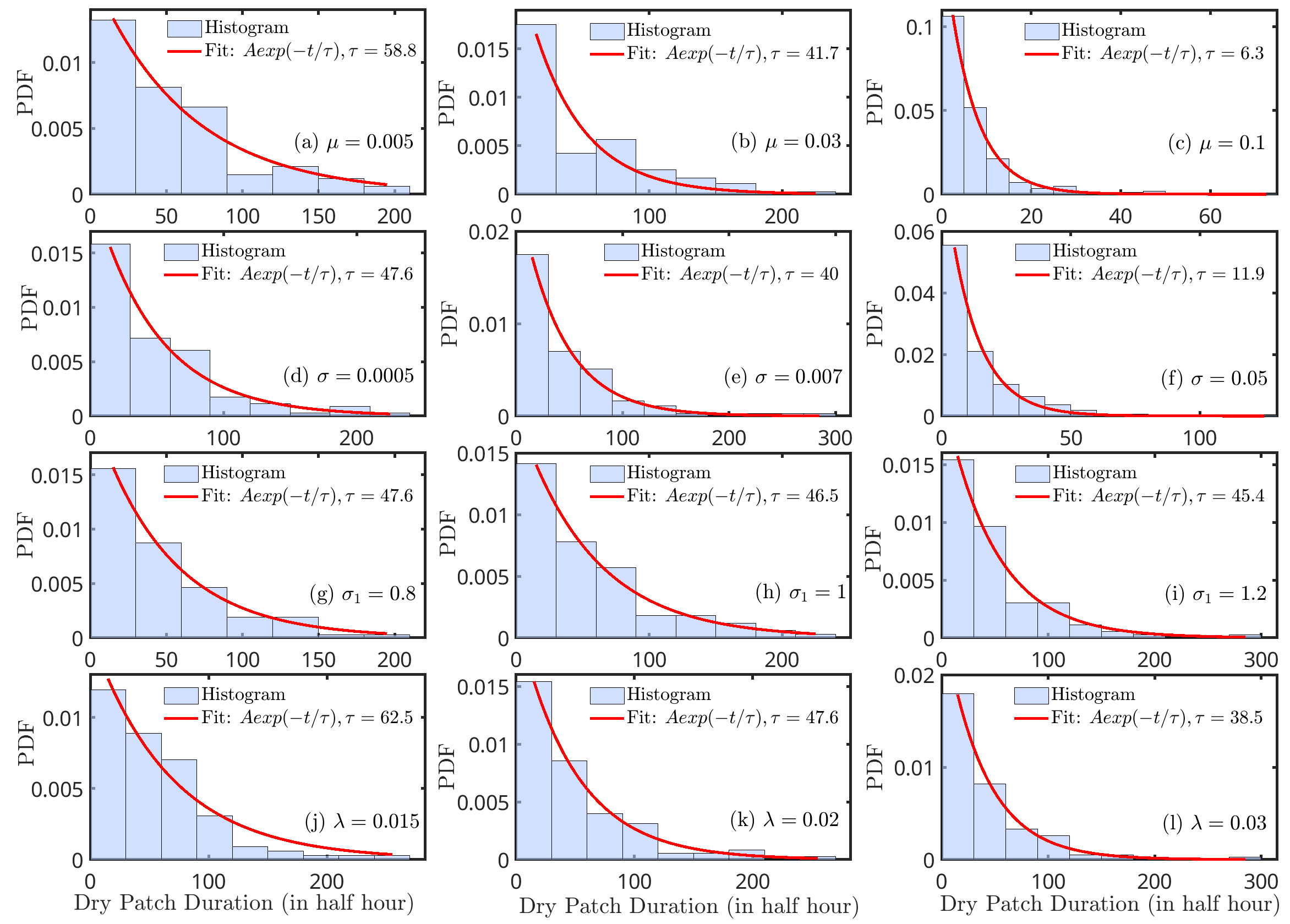}
    \caption{Variation of the PDF of dry patch duration for different parametric values with exponential fit (in red). Four different rows indicate the gradual increment in the parametric values of $\mu$ ((a)-(c)), $\sigma$ ((d)-(f)), $\sigma_{1}$ ((g)-(i)) and $\lambda$ ((j)-(l)) respectively. $\tau$ indicates the characteristic dry patch duration for each case.}
    \label{fig6}
\end{figure*}

In Fig.~\ref{fig5}, we depict the impact of the variation in the $\sigma_{1}$ and $\lambda$ on the extreme events present in rainfall data obtained from the simulation of the stochastic model. Here the lower x-axis (shown with blue color) represents the values of $\sigma_{1}$ and the upper x-axis (shown with the red color) shows the variation with the $\lambda$ in each sub-figures. In Fig.~\ref{fig5}(a), we show the highest rainfall amplitude obtained for different values of $\sigma_{1}$ (in blue) and $\lambda$ (in red). With increasing values of $\sigma_{1}$ and $\lambda$, the average highest rainfall amplitude (averaged over twenty realizations) increases. With increasing value of $\sigma_{1}$, the fluctuation in the jump amplitude of the arrival process increases and hence more high intensity rain occurs [see~\ref{app_fig1}((j)-(l))]. $\lambda$ defines the mean intensity of the arrival process and, with increasing $\lambda$, a greater number of rainfall events occurs that increases the amplitude of rainfall [see~\ref{app_fig1}((m)-(o))]. In Fig.~\ref{fig5}(b), we show the averaged number of extreme events (averaged over twenty realizations) present in the simulated rainfall time series obtained for different values of $\sigma_{1}$ and $\lambda$. With increasing values of both of them, the number of extreme events gradually increases. Here we can see that the number of extreme events highly depends on the value of $\lambda$ as it controls the arrival process term in the model. In Fig.~\ref{fig5}(c), we show the average percentage of rain coming from extreme events (averaged over twenty realizations). here we see that with increasing $\sigma_{1}$ values, averaged percentage of rain through extreme events increases rapidly as higher $\sigma_{1}$ values indicate more high amplitudes in rainfall time series [see~\ref{app_fig1}((j)-(l))]. Whereas $\lambda$ controls the intensity of the arrival process which increases overall rain in the rainfall time series with increasing $\lambda$ and thus, the percentage of rainfall through extreme events does not depend much on the values of $\lambda$.

The analysis on extreme events reveal its dependence on different parameters of our stochastic model. There are several studies on the extreme events present in the rainfall time series, specially in North-East region of India~\cite{singh2024trends,saha2023present,das2025changing,sharma2025seasonal}. The reported trend of decreasing heavy rainfall while intense extreme events and longer dry patch duration indicate towards the notion `wet regions are becoming wetter, while dry regions are becoming drier'~\cite{singh2024trends}. In the next sub-section we explain our findings regarding dry patch duration of the simulated rainfall time series.

 \subsubsection{Statistics of dry patch}

A dry patch is defined as a contiguous interval in the rainfall time series during which the precipitation amplitude is effectively zero. In practice, rainfall $\leq 0.1$ mm per half-hour is treated as zero, and at least two consecutive such intervals are required to identify a dry patch. This criterion removes spurious fluctuations and ensures physically meaningful rain-free periods.

We consider only those simulated time series whose total accumulated rainfall lies within $5000 \pm 500$ mm over six months, ensuring comparability across parameter sets. Figure~\ref{fig6} shows the probability density functions (PDFs) of dry patch durations for different parameters, along with exponential fits (red curves). The extracted scale $\tau$ characterizes the typical persistence time of dry intervals.

With increasing mean rainfall intensity $\mu$ [Figs.~\ref{fig6}(a)–(c)], the system shifts toward a wetter regime, leading to a rapid suppression of dry periods. Accordingly, $\tau$ decreases sharply from $58.8$ to $6.3$ half-hour units. A similar trend is observed with increasing variability $\sigma$ [Figs.~\ref{fig6}(d)–(f)], where enhanced fluctuations promote intermittent rainfall events that fragment dry intervals, reducing $\tau$ from $47.6$ to $11.9$ half hours.

In contrast, increasing the amplitude variability of the arrival process $\sigma_{1}$  has only a weak effect, with $\tau$ decreasing marginally from $47.6$ to $45.4$ half hours as shown in Figs.~\ref{fig6}(g)–(i). This indicates that variability in event magnitude alone does not significantly alter the temporal organization of dry periods. However, increasing the mean arrival rate $\lambda$ [Figs.~\ref{fig6}(j)–(l)] strongly suppresses dry intervals by increasing event frequency, leading to a reduction in $\tau$ from $62.5$ to $38.5$ half hours.

\begin{figure*}
    \centering
     \includegraphics[width=1\textwidth]{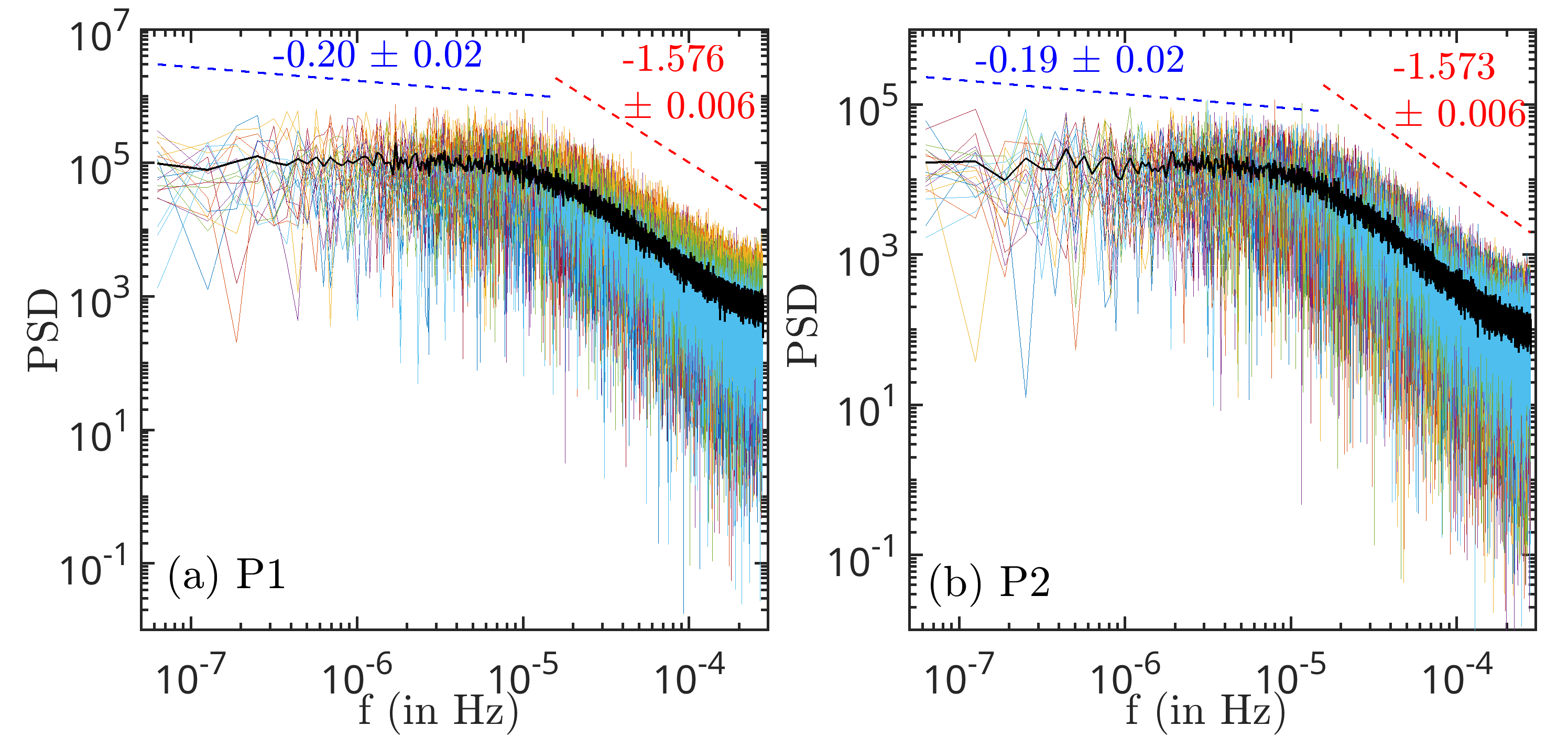}
    \caption{Power spectral density (PSD) of simulated rainfall time series. The thin colored lines correspond to twenty different realizations and the thick black line shows their mean behavior. (a) PSD of the simulated rainfall data which follows the Log-Normal PDF (b) PSD of the simulated rainfall data which follows the Gamma PDF.}
    \label{fig7}
\end{figure*}

The exponential form of the PDFs suggests that the termination of dry patches is approximately memory-less, consistent with a Poisson-like arrival of rainfall events. The dependence of $\tau$ on model parameters thus provides a direct physical measure of how rainfall intensity, variability, and event frequency control the persistence of dry conditions.

In extreme event analysis, we find the maximum rain intensity as well as number of extreme events increase with increasing $\lambda$ and $\sigma_{1}$ values. Whereas with increasing values of these two parameters, the characteristic length of the dry patch duration decrease gradually, indicating shorter dry patch and more and intense extreme events may present for higher values of these parameters. We need more observational data to verify this predicted characteristic. However, this finding is in line with the reported prediction over the nature of rainfall due to climate change\cite{singh2024trends}.  

\subsection{Spectral analysis of simulated rainfall data}

To further explore the temporal characteristics of the simulated rainfall time series, we perform a spectral analysis. This analysis provides insight into the distribution of variance across different frequencies, allowing us to identify dominant time scales and the influence of extreme events on rainfall dynamics. We approach this in two complementary ways: first, through power spectrum analysis to examine global scaling behavior and frequency-domain correlations, and second, using wavelet analysis combined with intrinsic mode functions to capture local, time-dependent variations and multi-scale features of the rainfall series. Together, these methods offer a comprehensive view of both persistent patterns and intermittent fluctuations inherent in the simulated rainfall data.
\subsubsection{Power spectrum analysis}


The power spectrum provides a quantitative view of the dominant frequencies present in a time series and offers insight into the type of noise and correlations inherent in the system. For rainfall data, the power spectral density (PSD) can reveal both short-term fluctuations and long-term temporal correlations, including the signature of extreme events. To explore these features, we compute the PSD of the simulated rainfall time series over multiple realizations. In our analysis, twenty independent realizations of the simulated series are considered, with the thin colored lines representing individual realizations and the thick black curve showing the averaged PSD.

\begin{figure*}[!htp]
    \centering
        \subfloat{\includegraphics[width=0.48\textwidth]{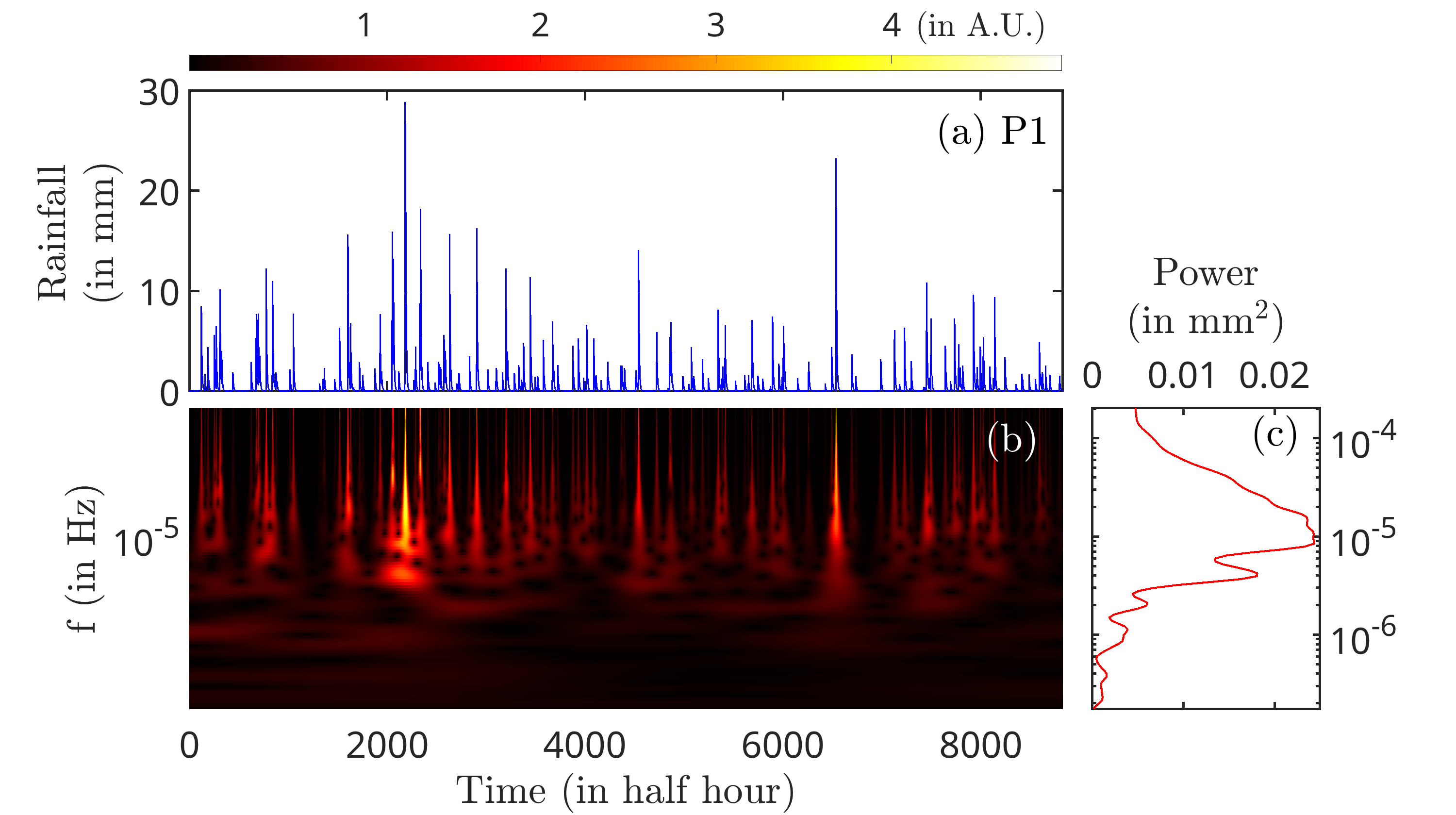}}
        \subfloat{\includegraphics[width=0.5\textwidth]{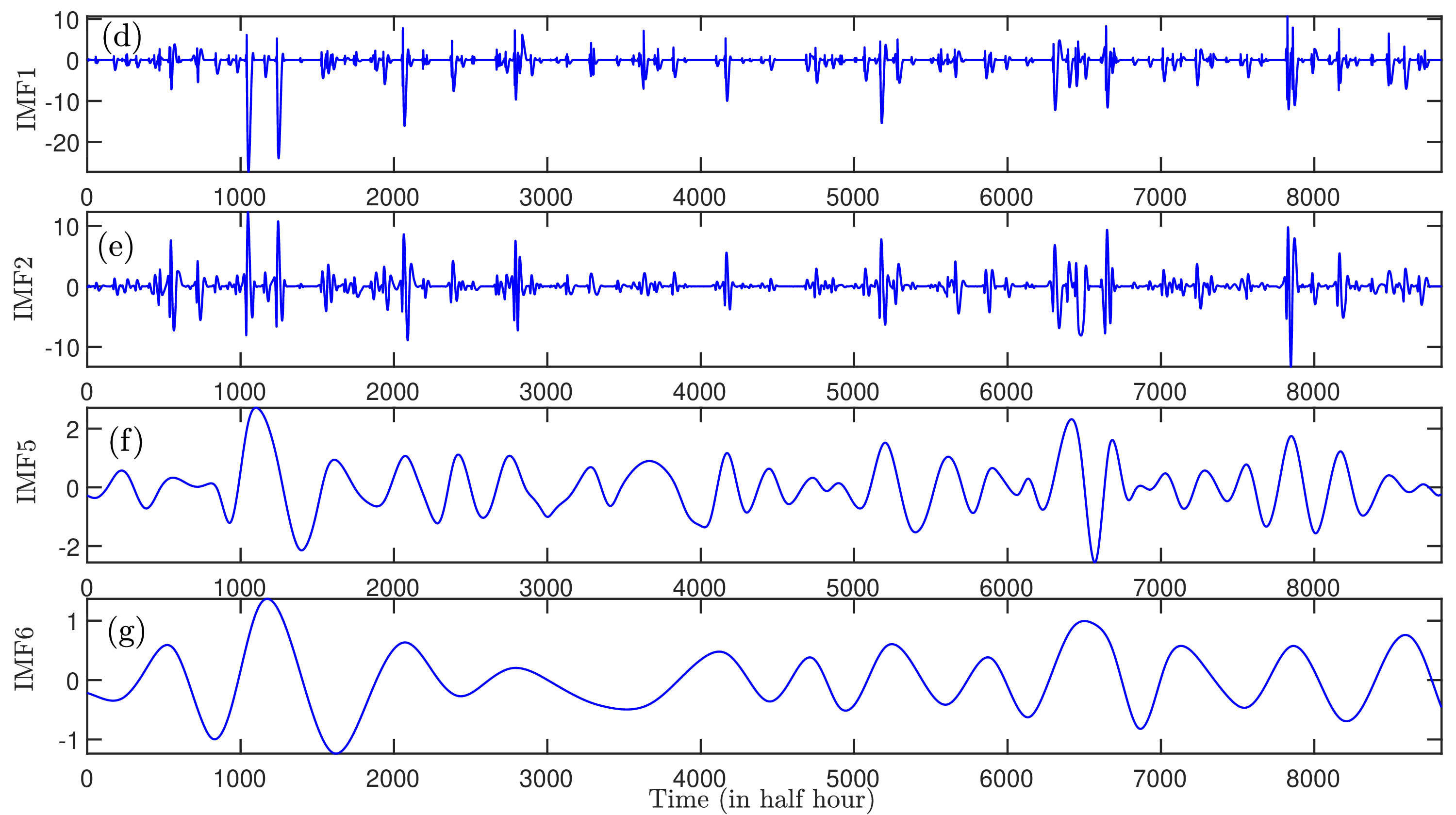}}\\
        \subfloat{\includegraphics[width=0.48\textwidth]{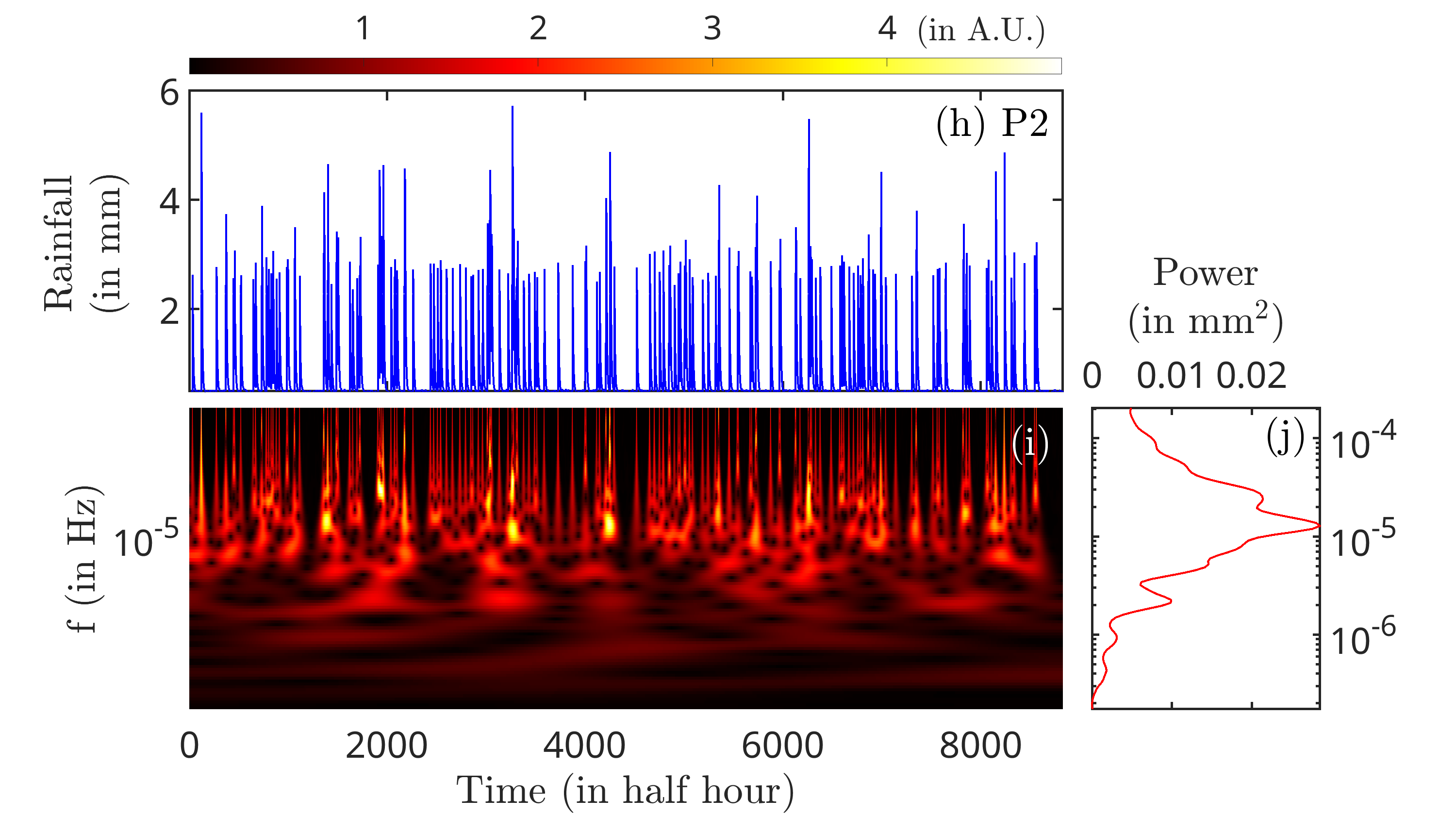}}
        \subfloat{\includegraphics[width=0.5\textwidth]{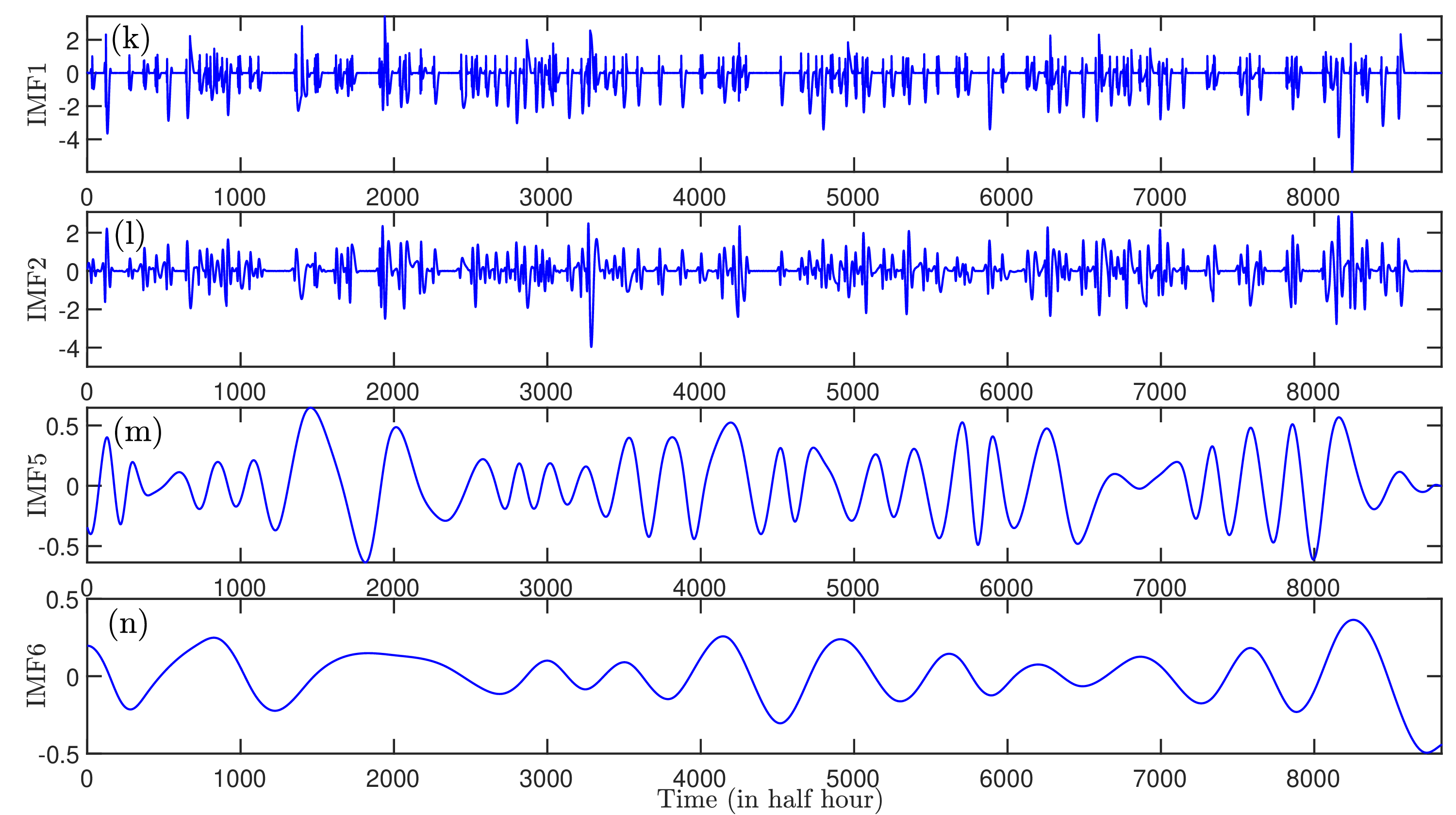}}
    \caption{Wavelet analysis of the simulated rainfall time series. (a)Temporal evolution of the simulated rain data. (b)Wavelet spectrum of the data shown in (a). (c)Global wavelet power spectrum of the same data showing the local frequency peaks present in the time series.}
    \label{fig8}
\end{figure*}

For the parameter set P1, corresponding to rainfall amplitudes following a Log-Normal distribution, the simulated series contains more extreme events compared to series following a Gamma distribution. The averaged PSD exhibits clear power-law behavior, with distinct scaling in the high- and low-frequency regions. In the high-frequency range, the PSD follows a power law with an exponent of $-1.576 \pm 0.006$, while in the low-frequency range, the exponent is $-0.20 \pm 0.02$. These values are in good agreement with the exponents observed in empirical rainfall data, which are approximately $-1.5$ and $-0.3$, respectively. Similar high-frequency scaling behavior has been reported in previous studies of rainfall spectra.

For the parameter set P2, where rainfall amplitudes follow a Gamma distribution, the PSD also exhibits power-law scaling in both frequency regions. The high-frequency exponent remains nearly unchanged ($-1.573 \pm 0.006$), while the low-frequency exponent is slightly smaller ($-0.19 \pm 0.02$) due to the lower prevalence of extreme rainfall events in the series. These results indicate that the low-frequency scaling of the PSD is sensitive to the presence of extreme events, while the high-frequency behavior is largely insensitive to the underlying distribution. Thus, the analysis of the PSD provides a clear means to characterize differences in rainfall dynamics associated with varying extreme-event statistics.

\subsubsection{Wavelet analysis and Intrinsic mode functions}

\begin{figure*}[!htp]
    \centering
        \includegraphics[width=1\textwidth]{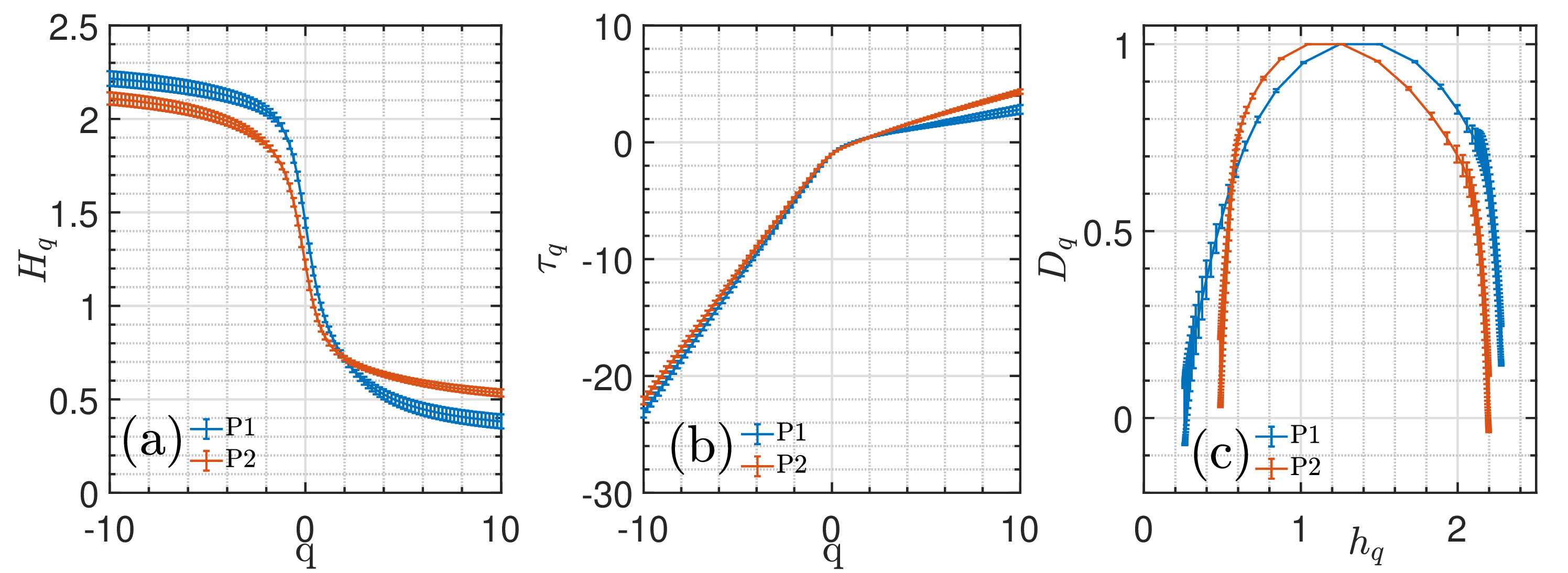}
    \caption{Multifractal analysis of the simulated rainfall time series. Blue curves are obtained by analyzing the simulated rainfall data which follows Log-Normal PDF and orange curves are obtained using the simulated rainfall data which follows Gamma PDF. (a) Variation of generalized Hurst exponent ($H_{q}$) with different moments ($q$). (b) Variation of mass exponent ($\tau_{q}$) with different moments ($q$). (c) Variation of multifractral spectrum ($D_q$) with singularity exponent (or H$\ddot{o}$lder exponent ($h_{q}$)).  }
    \label{fig9}
\end{figure*}

Rainfall is inherently a multiscale phenomenon, exhibiting dynamics across a wide range of local time scales. To probe these temporal features, we employ wavelet analysis~\cite{PhysRevE.58.2779,torrence1998practical}, which is particularly well-suited for capturing localized variations in frequency and amplitude. Unlike the traditional Fourier-based power spectrum that relies on continuous sine and cosine functions, wavelet analysis uses predefined wavelets of various shapes as basis functions, enabling a detailed examination of non-stationary and intermittent signals such as rainfall. In our model, parameters extracted from observed rainfall data of the North-East region of India, particularly the mean rainfall ($\mu$) and the mean intensity of the arrival process ($\lambda$), allow us to incorporate local periodicities into the simulated series.

For the parameter set P1, the wavelet power spectrum of the simulated rainfall time series reveals both time- and frequency-dependent variations. Here, the x-axis represents time, the y-axis represents frequency, and the color scale denotes power. The corresponding global wavelet power spectrum, obtained by averaging the wavelet spectrum over time, captures the dominant local frequencies present in the series. Using this analysis, we identify characteristic time scales of approximately 21.2 days, 2.7 days, 18.7 hours, and 16.6 hours, which are in good agreement with those found from observed rainfall data (21 days, 4 days, 24 hours, and 12 hours, respectively)~\cite{ghoshdastider2025kolmogorov}.

To further examine these local time scales, we perform Empirical Mode Decomposition (EMD) to extract intrinsic mode functions (IMFs)~\cite{huang1996mechanism,huang1998empirical,huang1999new}. These IMFs represent sub-signals within the original series, obtained empirically without any preassigned filter. This data-driven decomposition isolates the local time scales identified in the wavelet analysis. For the parameter set P1, the extracted IMFs confirm the presence of these characteristic temporal features, validating the consistency of the simulated data with observed rainfall dynamics.

For the parameter set P2, corresponding to rainfall amplitudes following a Gamma distribution, the wavelet analysis similarly reveals local time scales of approximately 26 days, 3 days, 19 hours, and 13.3 hours. Compared with P1 and the observed data, the main difference arises in the longer time scales (low-frequency components), reflecting the reduced number of extreme events in the Gamma-dominated rainfall series. The corresponding IMFs for P2 capture these local scales and further confirm the variation in temporal structure induced by changes in the underlying statistical distribution.

Overall, the combination of wavelet analysis and EMD provides a comprehensive view of both global and local temporal dynamics in simulated rainfall time series, capturing the multiscale structure and the influence of extreme events across different parameter regimes.
\subsection{Multifractal analysis of simulated rainfall data}

The multifractal behavior of the rainfall time series is a well studied phenomenon. The power law behavior of the PSD and the existence of a long tail in PDF indicates towards the possible multifractal behavior of the rainfall data. We have analyzed the multifractal features of the observed rainfall time series of North-East region of India~\cite{ghoshdastider2025kolmogorov}. In this section, we performed multifractal analysis to our simulated rainfall time series obtained using Eq.~(\ref{eq1}). It is an essential tool to investigate how the scalings are changing with different moments of highly fluctuating rainfall time series. For that we employed multifractal detrended fluctuation analysis (MFDFA) \cite{ihlen2012introduction,sarker2021detrended,agbazo2019fractal} to study how fluctuations present in the rainfall data behave at different moments ($q$) with different length scale of the rainfall time series. Multifractal spectrum obtained from this analysis provides insights into the different degree of complexity and the dominance of large and small fluctuations in the rainfall process. 

After analysis, we can show that for multifractal systems, the fluctuation function ($F_{q}(s)$) shows a power law behavior with different segment sections of the time series having length $s$ such that

\begin{equation}
F_{q}(s) \sim s^{H_{q}} \label{eq6}
\end{equation}

for any fixed value of $q$. This exponent $H_{q}$ is called the generalized Hurst exponent. Using it, mass exponent ($\tau_{q}$) can be calculated by the following equation:

\begin{equation}
\tau_{q}= q*H_{q} - 1 \label{eq7}
\end{equation}

The multifractal spectrum ($D_{q}$) of $q$th order can be obtained through the following equations:
we need  which is related to $\tau_{q}$ via the Legendre transformation such that 
\begin{equation}
    \begin{aligned}
      h_{q}= \tau^{\prime}_{q} \\
      D_{q}=(q\times h_{q}) - \tau_{q}  \label{eq8}
    \end{aligned}
\end{equation}
 where $h_{q}$ is known as the singularity strength or as H$\ddot{o}$lder exponent.

In Fig.~\ref{fig9}, we show the multifractal analysis of the simulated rainfall time series where the blue curves refer to the simulated rainfall time series obtained for the parameter set P1 and the orange curves show the simulated rainfall time series obtained for the parameter set P2. The generalized Hurst exponent ($H_{q}$) does not depend on different order of moments ($q$) for the monofractal time series but in the multifractal case it varies with $q$. In Fig.~\ref{fig9} (a), we show the averaged variation of the generalized Hurst exponent ($H_{q}$) with different moments ($q$) (averaged over twenty different realizations obtained by simulating Eqn.~(\ref{eq1})). Here we see that $H_{q}$ varies with different values of $q$ which is a signature of multifractality. For the negative values of $q$, $H_{q}$ describes the behavior of the segments with small fluctuations. On the other hand, for the positive values of $q$, $H_{q}$ depicts segments with large fluctuations. In Fig.~\ref{fig9} (a), we see that the behavior of $H_{q}$ is almost the same for the lower order of moments (for both positive and negative moments) for both of the curves for P1 and P2, while they separate at the higher ends for higher values of moments (for both positive and negative moments). The presence of more extreme events and overall more local fluctuations in the time series of rainfall following Log-Normal PDF (the parameter set P1) contributes to lower values of $H_{q}$ at positive higher moments and higher $H_{q}$ values at negative higher moments.

The Hurst exponent ($H$) is defined as the $H_{q}$ value for $q=2$ and its value indicates whether the time series under investigation is persistent or not and its predictability. The mean value of $H$ averaged over twenty different realizations is found to be close to $0.73$, indicating the simulated rainfall time series is persistent in nature and thus carry the predictable features for a shorter time period by knowing the initial state. For real observed rainfall data of North-East region of India, mean Hurst exponent comes out to be $0.64$~\cite{ghoshdastider2025kolmogorov}. So, the value obtained using the simulated data is considerably close to the observed value (within $14\%$).

In Fig.~\ref{fig9} (b), we show the variation of mass exponent ($\tau_{q}$) with different order of moments ($q$) for the simulated rainfall time series obtained for the parameter set P1 (in blue) and the same obtained for the parameter set P2 (in orange). Here we see that the mass exponent varies non-linearly with $q$ which is the signature of multifractality present in the time series. From the relation between $H_{q}$ and $\tau_{q}$, we can see for higher positive moments, $\tau_{q}$ has higher values for the Gamma PDF dominated simulated rainfall data (parameter set P2) than the Log-Normal dominated simulated rainfall data (parameter set P2) due to the presence of more extreme events in the latter. 

The multifractal spectrum carries information about the distribution of the scaling exponents of the multiscale systems at different time scales. In Fig.~\ref{fig9} (c), we show the multifractral spectrum ($D_q$) for the simulated rainfall time series that follows Log-Normal PDF (in blue) and the same that follows Gamma PDF (in orange) obtained by simulating Eqn.~(\ref{eq1}). We notice that the spectrum with respect to singularity exponent (or H$\ddot{o}$lder exponent $h_q$) exhibits a wide distribution; a typical characteristic of the multifractal nature of rainfall events. The width of the multifractal spectrum provides a direct measure of the degree of complexity in the multifractal nature of the system. In Fig.~\ref{fig9} (c), we observe that the spectrum width of the simulated data obtained for the parameter set P2 is lesser than that of the simulated rainfall data obtained for the parameter set P1. The left side of the spectrum corresponds to the presence of the extreme events in the simulated data, while the right side of the spectrum indicates the extent of local fluctuations in the same. The asymmetric form of the spectrum of simulated rainfall data for the parameter set P2 (orange) indicates an asymmetric distribution of local and high fluctuations in the data which is a distinct feature compared to that of the simulated rainfall data obtained using the parameter set P1.

\section{Summary and Conclusions}\label{sec5}

In this work, we have developed a stochastic jump-diffusion model to generate synthetic rainfall time series and validated it using half-hourly observational data from the North-East region of India over a twenty-year period. The model provides a flexible framework to investigate the influence of different parameters on rainfall dynamics, allowing us to probe both typical and extreme events systematically.

Our analysis demonstrates that the simulated rainfall time series captures the intermittent nature of the rainfall process. The time-averaged mean square displacement of the simulated data exhibits superdiffusive behavior, closely matching the observed data, indicating that the model effectively reproduces the underlying stochastic dynamics. The probability distribution of rainfall intensity in the simulated series, when extracted from observed data parameters, follows a Log-Normal form, consistent with empirical observations. By varying key parameters, we show that both the mean rainfall and the standard deviation of the arrival process amplitude are critical in determining whether the simulated data follow a Log-Normal or Gamma distribution. Mapping the parameter space reveals distinct regions where the simulated rainfall transitions between these two characteristic distributions.

Extreme rainfall events are strongly influenced by the amplitude and intensity of the arrival process. Increasing these parameters leads to higher maximum rainfall intensities, a larger number of extreme events, and a greater fraction of total rainfall contributed by these extremes. Similarly, the structure of dry patches is sensitive to model parameters: higher mean rainfall, variability, and arrival intensity reduce the characteristic dry patch duration, reflecting a more continuous rainfall regime.

In the frequency domain, power spectral analysis shows that the simulated rainfall series exhibits power-law scaling, with distinct exponents in low- and high-frequency regions. Comparing two representative parameter sets, corresponding to Log-Normal and Gamma distributions, reveals that rainfall series dominated by Log-Normal behavior contain more extreme events, reflected in slightly steeper low-frequency exponents. Local temporal scales extracted using wavelet analysis and validated via empirical mode decomposition demonstrate that the model reproduces the dominant cycles observed in the empirical data, including daily and multi-day variability, with differences in extreme-event-rich series corresponding to the longest time scales.

Finally, multifractal analysis using the MFDFA method confirms that the simulated rainfall series captures the complex hierarchical structure of rainfall variability. Series generated under the Log-Normal-dominated parameter set display broader multifractal spectra, reflecting the presence of extreme events and stronger variability across scales, in agreement with observations. Comparisons of generalized Hurst exponents and mass exponents highlight that differences between the parameter sets are most pronounced for large fluctuations, emphasizing the role of model parameters in controlling the extremes and multifractal complexity of rainfall dynamics.

The present stochastic jump-diffusion framework provides a robust tool for reproducing rainfall statistics, including intermittent bursts, extreme events, and multifractal characteristics. There are several promising directions to extend the present stochastic rainfall model. One natural extension is to incorporate spatio-temporal correlations to study how extreme rainfall events propagate across regions and interact with larger-scale atmospheric variability, drawing parallels with correlated stochastic transport processes~\cite{dhar2008heat}. Another direction is to explore parameterized stochastic forcing to investigate the emergence of extreme events and multifractal scaling under slowly varying environmental conditions, akin to approaches used in anomalous diffusion and non-equilibrium fluctuation studies~\cite{Kundu_2011}.

\section{Acknowledgment}\label{sec6}

J. GhoshDastider acknowledges the Ministry of Education (MOE), Government of India, for providing financial support for her research work through the Prime Minister’s Research Fellowship (PMRF) May 2022 scheme.

\onecolumngrid
\appendix \label{app} 
\counterwithin{figure}{section}

\section{Numerical scheme}\label{app1}

In this Appendix, we outline the numerical scheme used to solve the proposed stochastic differential equation [Eq.~\ref{eq1}] describing rainfall dynamics. We begin with a general stochastic differential equation (SDE) of the form
\begin{equation}
dX_{t}= a(X_{t},t)dt + b(X_{t},t)dW_{t},
\label{eqn1}
\end{equation}
where $a(X_t,t)$ and $b(X_t,t)$ denote the drift and diffusion coefficients, respectively, and $W_t$ is a standard Wiener process satisfying $E[W_t]=0$ and $E[W_{t_1}W_{t_2}]=\min(t_1,t_2)$.

For a sufficiently smooth function $f(X_t,t)$, Ito's lemma gives
\begin{equation}
df = \frac{\partial f}{\partial t}dt + a(X_t,t)\frac{\partial f}{\partial x}dt
+ \frac{b^2(X_t,t)}{2}\frac{\partial^2 f}{\partial x^2}dt
+ b(X_t,t)\frac{\partial f}{\partial x},dW_t,
\label{eqn2}
\end{equation}
with the standard Ito rules: $dt^2=0$, $dt dW_t = dW_t dt = 0$, and $(dW_t)^2 = dt$.

In integral form, this can be expressed as
\begin{equation}
f(X_t,t)=f(X_0,t_0)+\int_{t_0}^{t} L^{0}f(X_s,s)ds + \int_{t_0}^{t} L^{1}f(X_s,s)dW_s,
\label{eqn3}
\end{equation}
where the differential operators are defined as $L^{0}= \frac{\partial }{\partial t}+a\frac{\partial }{\partial x}+\frac{b^{2}\partial^{2}}{2 \partial x^{2}}$ and $L^{1}=b\frac{\partial }{\partial x}$.
Choosing $f(X_t,t)=X_t$, we obtain the integral representation of the SDE:
\begin{equation}
X_t = X_0 + \int_{t_0}^{t} a(X_s,s),ds + \int_{t_0}^{t} b(X_s,s),dW_s.
\label{eqn4}
\end{equation}

A stochastic Taylor expansion is constructed by recursively expanding the coefficients $a(X_s,s)$ and $b(X_s,s)$ using Eq.~\ref{eqn3}. Retaining terms up to first order yields
\begin{equation}
X_t \approx X_0 + a(X_0,t_0)(t-t_0) + b(X_0,t_0)(W_t - W_{t_0}),
\label{eqn6}
\end{equation}
which forms the basis of a first-order weak approximation.

Discretizing time as $t_n = t_0 + n\Delta t$, the above approximation leads to the Euler–Maruyama scheme:
\begin{equation}
X_{t_{n+1}} = X_{t_n} + a(X_{t_n},t_n)\Delta t + b(X_{t_n},t_n)\Delta W_{t_n},
\label{eqn7}
\end{equation}
where the Wiener increment is given by $\Delta W_{t_{n}} \approx N(0,1) \sqrt(\Delta t_{n})$,i.e., $\Delta W_{t_n}$ is a Gaussian random variable with zero mean and variance $\Delta t$, which can be generated numerically as $\sqrt{\Delta t},N(0,1)$.

This scheme naturally separates the rainfall dynamics into three physically distinct components: (i) the drift term governing the baseline rainfall evolution, (ii) the diffusion term capturing continuous stochastic fluctuations, and (iii) the jump term representing discrete, burst-like precipitation events. The interplay of these contributions enables the model to reproduce both gradual variability and intermittent extreme rainfall typical of observed precipitation time series.
\section{Simulated rainfall time series in different parametric region}\label{app2} 

\begin{figure*}[!htp]
    \centering
        \includegraphics[width=1\textwidth]{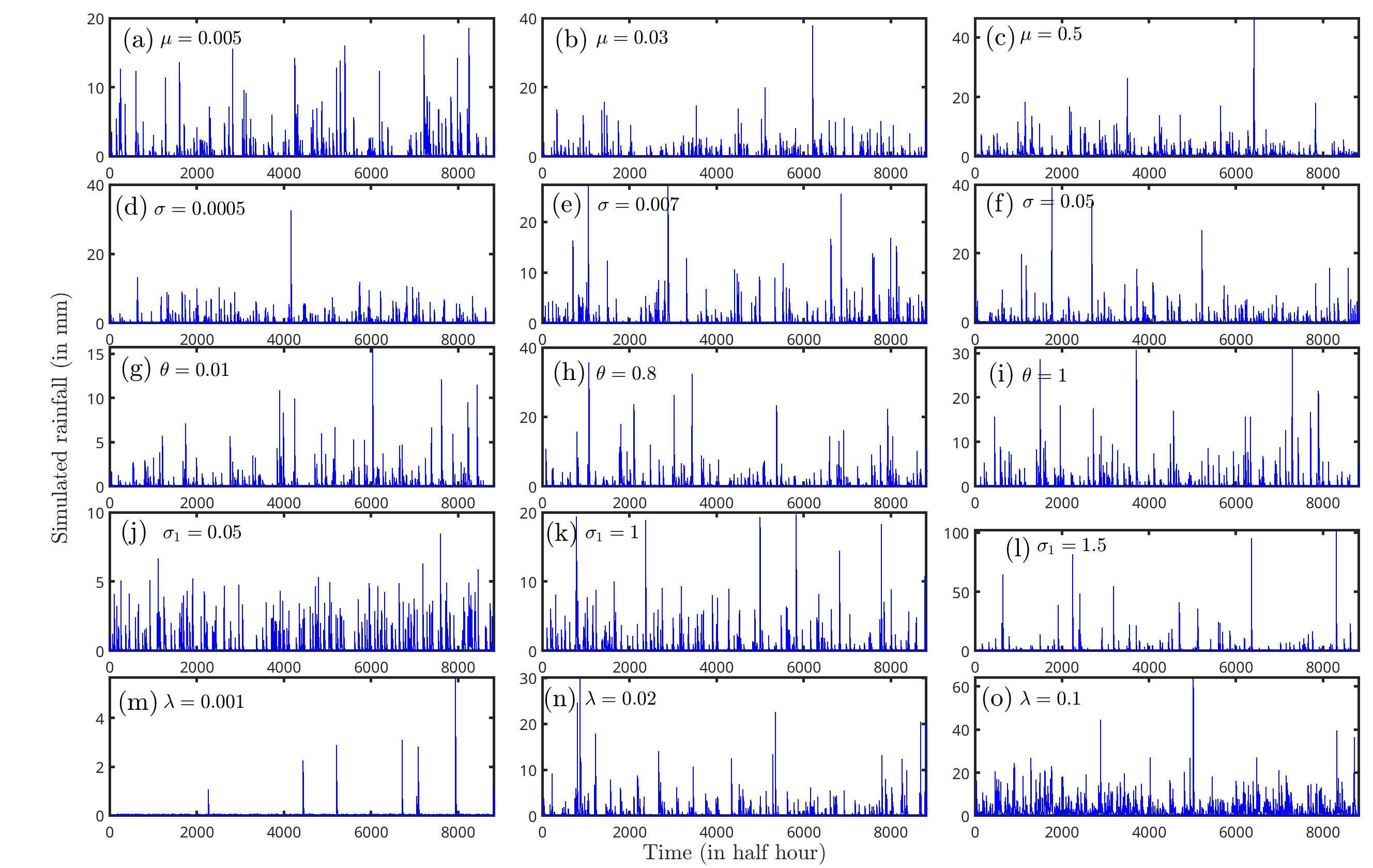}
    \caption{Temporal evolution of simulated rainfall data in different regions of different parameters. Five different rows contain the simulated rainfall data for five different parameters and from left to right, values of the parameters increases gradually. }
    \label{app_fig1}
\end{figure*}

In this Appendix, we illustrate how the statistical properties of the simulated rainfall time series depend on key model parameters: the mean rainfall $\mu$, its standard deviation $\sigma$, the mean jump amplitude $\theta$, the amplitude variability $\sigma_{1}$, and the mean arrival rate $\lambda$.

Figure~\ref{app_fig1} shows representative temporal evolutions generated from Eq.~(\ref{eq1}) under systematic variation of these parameters. Panels (a)–(c) correspond to increasing $\mu$, which sets the baseline rainfall level. As $\mu$ increases, the time series shifts toward a uniformly wetter regime, with higher cumulative rainfall primarily arising from sustained moderate precipitation rather than isolated extremes.

Panels (d)–(f) show the effect of increasing $\sigma$, which controls the strength of fluctuations about the mean. Larger $\sigma$ leads to more pronounced variability, with intermittent high-intensity events superimposed on low-rain intervals. This enhanced variability results in a gradual increase in total rainfall, driven by sporadic bursts.

In panels (g)–(i), we vary $\theta$, the mean amplitude of the jump process $J_t$. Increasing $\theta$ raises the typical magnitude of rainfall events, thereby increasing overall precipitation through stronger individual bursts while preserving the temporal structure of arrivals.

Panels (j)–(l) illustrate the role of $\sigma_{1}$, the standard deviation of the jump amplitude. For small $\sigma_{1}$, rainfall events are relatively uniform and weak. As $\sigma_{1}$ increases, the distribution of event magnitudes broadens, leading to occasional high-intensity bursts and enhanced intermittency in the time series.

Finally, panels (m)–(o) show the effect of increasing the arrival rate $\lambda$ of the counting process $N_t$. At low $\lambda$, rainfall events are sparse and well separated. Increasing $\lambda$ results in more frequent arrivals, producing a denser sequence of rainfall bursts and a corresponding increase in cumulative rainfall.

Overall, these parameter variations highlight the distinct physical roles of baseline intensity ($\mu$), fluctuations ($\sigma$), event amplitude ($\theta$, $\sigma_{1}$), and event frequency ($\lambda$) in shaping both the magnitude and temporal structure of the rainfall process.
\section{Extreme events analysis of simulated rainfall time series}\label{app3}

\begin{figure*}[!htp]
    \centering
        \includegraphics[width=1\textwidth]{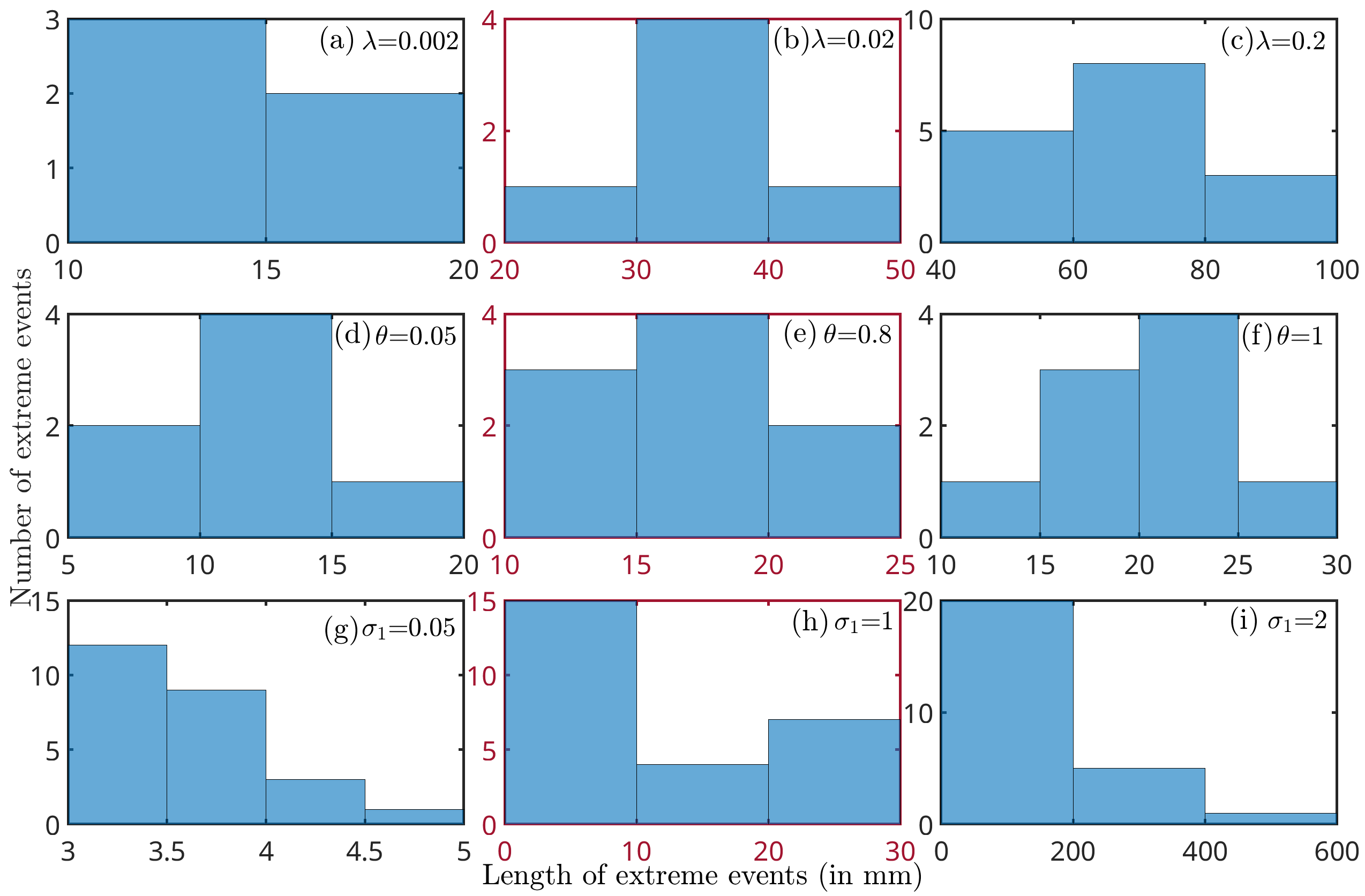}
    \caption{Variation of the PDF of extreme events of the simulated rainfall time series with different parametric values. Three rows are representing three different parametric values of the mean intensity of the arrival process ($\lambda$), the mean of the amplitude of the arrival process ($\theta$) and standard deviation of the amplitude of the arrival process ($\sigma_{1}$). }
    \label{app_fig2}
\end{figure*}

In this section, we discuss our analysis on the extreme events present in the simulated rainfall data in addition to the discussion in the main body of the paper. In Fig.~\ref{app_fig2}, we show the histogram of extreme events, where the x-axis represents the length or amplitude of extreme events (in mm) and the y-axis represents the number of extreme events present in the rainfall time series. In Fig.~\ref{app_fig2} (a)-(c), we show the evolution of the histogram of extreme events with increasing mean intensity of the arrival process ($\lambda$). In this case, we observe that the length (amplitude) of the extreme events as well as the number of extreme events increases gradually with increasing $\lambda$ values as higher value of $\lambda$ indicates higher number of rain events. In Fig.~\ref{app_fig2} (d)-(f), we show the evolution of the histogram of extreme events with increasing mean of the amplitude of the arrival process ($\theta$). With increasing $\theta$ values, the amplitude of the extreme events increases slowly while the number of extreme events are almost same.  In Fig.~\ref{app_fig2} (g)-(i), we show the evolution of the histogram of extreme events with increasing standard deviation of the amplitude of the arrival process ($\sigma_{1}$). The length of the extreme events increases drastically with increasing $\sigma_{1}$ values while the number of the extreme events increases slightly. 

\section{Multifractal analysis of simulated rainfall time series with varied parameters}\label{app4}
In this Appendix, we examine how the multifractal properties of the simulated rainfall time series depend on key model parameters. The analysis is carried out in terms of the generalized Hurst exponent $H_q$ and the corresponding multifractal spectrum, which together characterize the scaling behavior of fluctuations across different moments $q$.

\begin{figure*}[!htp]
    \centering
        \includegraphics[width=1\textwidth]{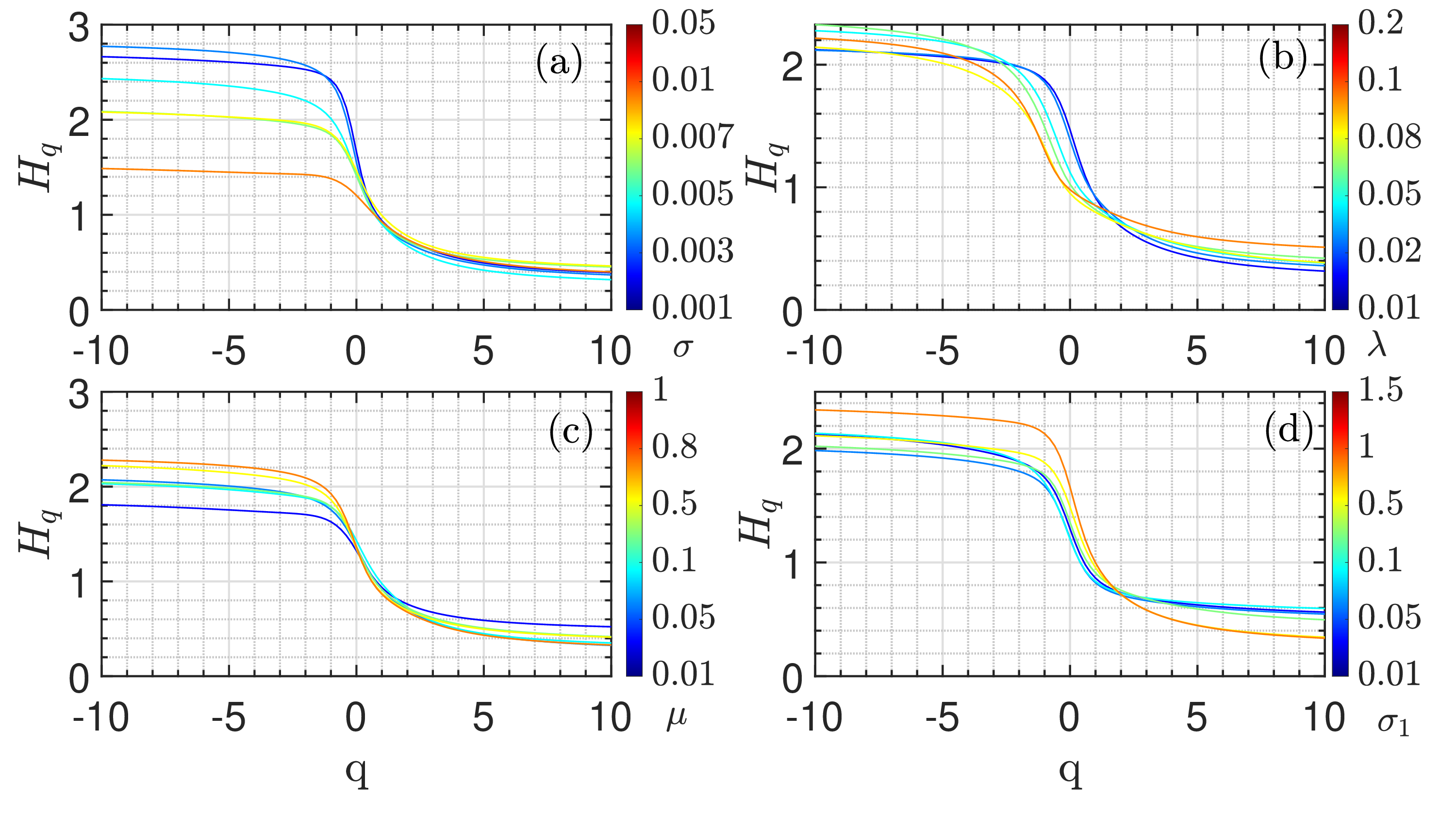}
    \caption{Variation of Generalized Hurst exponent $H_{q}$ with different moments ($q$) of the simulated rainfall time series. (a) to (d) shows the behaviour of it with the variation of $\sigma$, $\lambda$, $\theta$ and $\sigma_{1}$.}
    \label{app_fig4}
\end{figure*}

\begin{figure*}[!htp]
    \centering
        \includegraphics[width=1\textwidth]{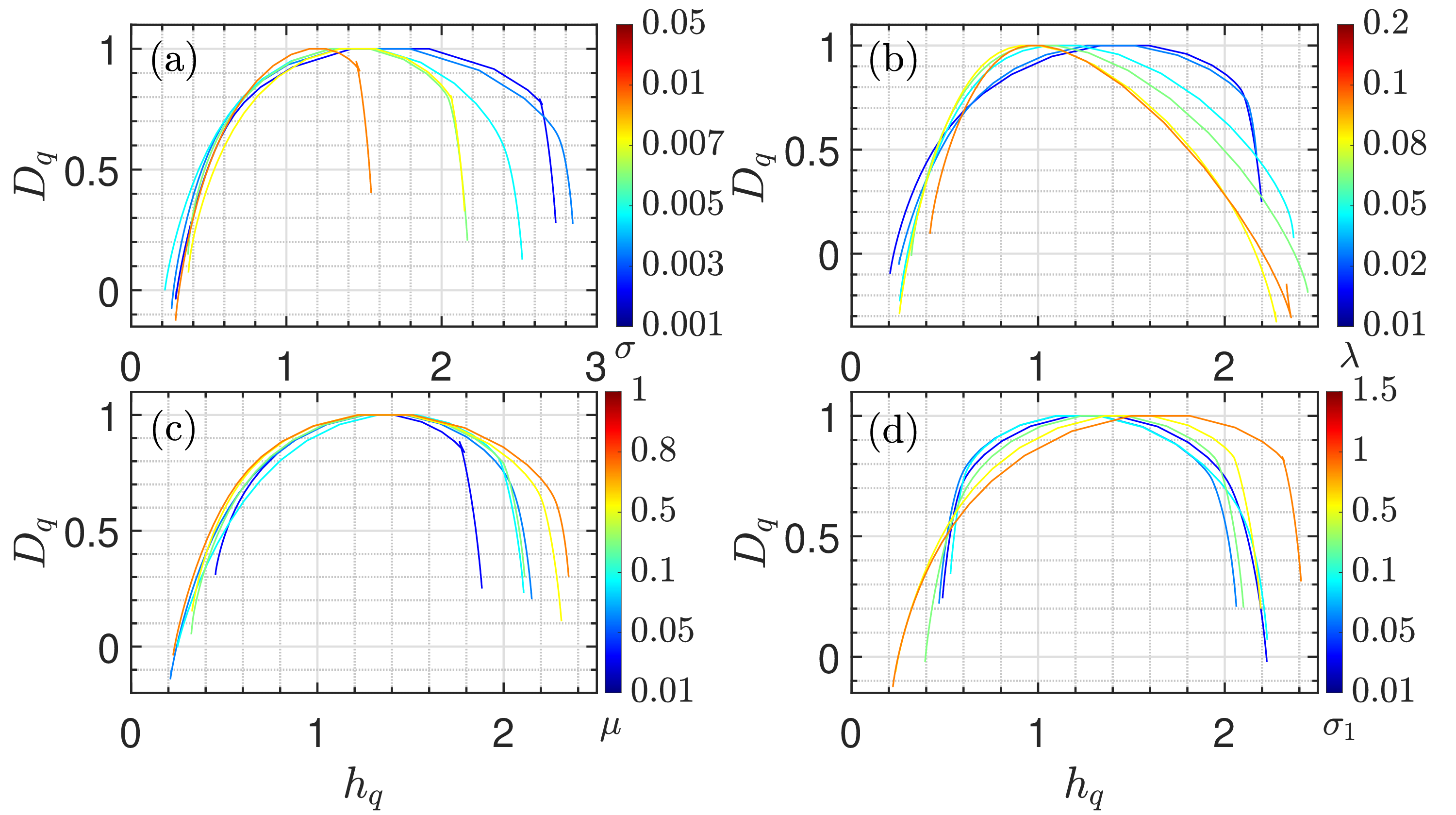}
    \caption{Variation of multifractral spectrum ($D_q$) with singularity exponent (or H$\ddot{o}$lder exponent ($h_{q}$)) of the simulated rainfall time series. (a) to (d) shows the behaviour of it with the variation of $\sigma$, $\lambda$, $\theta$ and $\sigma_{1}$.}
    \label{app_fig5}
\end{figure*}
Figure~\ref{app_fig4} shows the variation of $H_q$ with $q$ for different parameter values. In Fig.~\ref{app_fig4}(a), with increasing $\sigma$, the values of $H_q$ in the negative $q$ region decrease gradually, indicating a weakening of correlations associated with small fluctuations. This suggests that stronger variability suppresses the persistence of low-intensity rainfall events.

In Fig.~\ref{app_fig4}(b), variation in the arrival rate $\lambda$ primarily affects $H_q$ near small values of $q$, implying that $\lambda$ predominantly influences the scaling behavior of typical fluctuations rather than extreme events. In contrast, Fig.~\ref{app_fig4}(c) shows that changes in $\theta$ affect $H_q$ over a broader range of $q$ (both positive and negative), reflecting its role in modulating the overall amplitude of rainfall events.

Figure~\ref{app_fig4}(d) illustrates the effect of $\sigma_1$, where increasing $\sigma_1$ leads to a decrease in $H_q$ for positive $q$ and an increase for negative $q$. This indicates a redistribution of scaling behavior, with enhanced intermittency: large fluctuations become less correlated while small fluctuations gain relative prominence.

The corresponding multifractal spectra are shown in Fig.~\ref{app_fig5}. In Fig.~\ref{app_fig5}(a), increasing $\sigma$ leads to a gradual narrowing of the spectrum along with a loss of symmetry, indicating reduced multifractality and a dominance of irregular fluctuations. In Fig.~\ref{app_fig5}(b), increasing $\lambda$ induces a left-skewed asymmetry in the spectrum, signaling an increased contribution from large fluctuations due to more frequent rainfall events.

In Fig.~\ref{app_fig5}(c), increasing $\theta$ results in a broader multifractal spectrum, reflecting enhanced variability across scales and stronger local fluctuations. Finally, Fig.~\ref{app_fig5}(d) shows that increasing $\sigma_1$ significantly widens the spectrum, indicating a substantial increase in multifractality driven by the growing variability in event amplitudes.

Overall, these results demonstrate that different model parameters distinctly influence the scaling structure of rainfall: $\sigma$ and $\sigma_1$ primarily control intermittency and variability, $\lambda$ governs event frequency, and $\theta$ modulates event magnitude. Their combined effects determine the degree and nature of multifractality in the simulated rainfall time series.

\twocolumngrid
\bibliography{references}

\end{document}